\documentclass[traditabstract]{aa}

\usepackage{graphicx}
\usepackage{lscape}
\usepackage{amsmath}
\usepackage{txfonts}
\usepackage{url}
\usepackage{natbib}
\usepackage[usenames,dvipsnames,svgnames,table]{xcolor}
\usepackage{ulem}

\newcommand{\kms}{km~s$^{-1}$}
\newcommand{\Msun}{M$_\odot$}
\newcommand{\Rsun}{R$_\odot$}

\newcommand{\Mjup}{M$_J$}
\newcommand{\Rjup}{R$_J$}

\begin{document} 


\title{Hot Exoplanet Atmospheres Resolved\\ with Transit Spectroscopy (HEARTS)\thanks{Based on observations made at ESO 3.6~m telescope at the La Silla Observatory under ESO program 096.C-0331.}}
\subtitle{I. Detection of hot neutral sodium at high altitudes on WASP-49b}

\author{A.~Wyttenbach\inst{1}, C.~Lovis\inst{1}, D.~Ehrenreich\inst{1}, V.~Bourrier\inst{1}, L.~Pino\inst{1,2}, R.~Allart\inst{1}, N.~Astudillo-Defru\inst{1}, H.~M.~Cegla\inst{1,3}, K.~Heng\inst{4}, B.~Lavie\inst{1,4}, C.~Melo\inst{5}, F.~Murgas\inst{6,7}, A.~Santerne\inst{8}, D.~S\'egransan\inst{1}, S.~Udry\inst{1}, and F.~Pepe\inst{1}}
\authorrunning{Wyttenbach et al.}
\titlerunning{Transit Spectroscopy of WASP-49b}
\institute{Geneva Observatory, University of Geneva, ch. des Maillettes 51, CH-1290 Versoix, Switzerland\label{inst1}
\and Dipartimento di Fisica e Astronomia ``Galileo Galilei'', Universit\`a di Padova, Vicolo dell'Osservatorio 3, I-35122 Padova, Italy\label{inst2}
\and Astrophysics Research Centre, School of Mathematics \& Physics, Queen's University Belfast, University Road, Belfast BT7 1NN\label{inst3}
\and University of Bern, Center for Space and Habitability, Sidlerstrasse 5, CH-3012, Bern, Switzerland\label{inst4}
\and European Southern Observatory, Alonso de Cordova 3107, Vitacura Casilla 19001, Santiago 19, Chile\label{inst5}
\and Univ. Grenoble Alpes, IPAG, 38000 Grenoble, France\label{inst6}
\and CNRS, IPAG, 38000 Grenoble, France\label{inst7}
\and Aix Marseille Univ, CNRS, LAM, Laboratoire d'Astrophysique de Marseille, Marseille, France\label{inst8}\\
\email{aurelien.wyttenbach@unige.ch}}

\date{Received 2016-11-15; accepted 2017-01-23} 

\abstract{High-resolution optical spectroscopy during the transit of HD~189733b, a prototypical hot Jupiter, allowed the resolution of the \ion{Na}{i}~D sodium lines in the planet, giving access to the extreme conditions of the planet upper atmosphere. We have undertaken HEARTS, a spectroscopic survey of exoplanet upper atmospheres, to perform a comparative study of hot gas giants and determine how stellar irradiation affect them. Here, we report on  the first HEARTS observations of the hot Saturn-mass planet WASP-49b. We observed the planet with the HARPS high-resolution spectrograph at ESO 3.6m telescope. We collected 126 spectra of WASP-49, covering three transits of WASP-49b. We analyzed and modeled the planet transit spectrum, while paying particular attention to the treatment of potentially spurious signals of stellar origin. We spectrally resolve the \ion{Na}{i}~D lines in the planet atmosphere and show that these signatures  are unlikely to arise from stellar contamination. The large contrasts of $2.0\pm0.5\%$ (D$_2$) and $1.8\pm0.7\%$ (D$_1$) require the presence of hot neutral sodium ($2,950^{+400}_{-500}$~K) at high altitudes ($\sim$1.5~planet radius or $\sim$45,000~km). From estimating the cloudiness index of WASP-49b, we determine its atmosphere to be cloud free at the altitudes probed by the sodium lines. WASP-49b is close to the border of the evaporation desert and exhibits an enhanced thermospheric signature with respect to a farther-away planet such as HD~189733b.}

\keywords{Planetary Systems -- Planets and satellites: atmospheres, individual: WASP-49b -- Techniques: spectroscopic -- Instrumentation: spectrographs -- Methods: observational}

\maketitle

\section{Introduction}

There has been enormous progress in the characterization of exoplanets and their atmospheres. The most amenable systems for atmospheric studies are the transiting systems. During the transit, a minute fraction of the starlight is filtered by the planet atmospheric limb. At wavelengths where atmospheric components absorb this light, the partial occultation of the star by the planet (or transit depth) appears larger than in white light. Observing a transit with a spectrograph yields the transmission spectrum of the planet atmosphere \citep{Seager2000,Brown2001}. From this, one can constrain the temperature profile, composition, winds, and diffusion processes (Rayleigh scattering, clouds) in the atmosphere \citep[e.g.,][]{Madhusudhan2014}.

The signature of the resonant neutral sodium doublet (\ion{Na}{i} D) has proved to be a powerful probe of exoplanet atmospheres, especially with the medium-resolution ($\lambda/\Delta\lambda\sim5,500$) STIS spectrograph on board the {\it Hubble Space Telescope} \citep{Charbonneau2002,Sing2008a,Sing2008b,VidalMadjar2011b,VidalMadjar2011a,Huitson2012,Nikolov2014,Fischer2016,Sing2016}. Owing to its high cross section, the resonant \ion{Na}{i} doublet at 589~nm (Fraunhofer's D line) is extremely sensitive to small amounts of sodium, up to high altitudes in the planet atmosphere. The \ion{Na}{i} line cores, in particular, can trace the temperature profile up to the planet thermosphere, which has been measured effectively with the HARPS spectrograph on HD~189733b \citep{Wyttenbach2015, Heng2015}. Sodium is arguably the  easiest species to detect in a hot exoplanet atmosphere, and ground-based detections have been achieved for several gas giants \citep{Redfield2008,Snellen2008,Wood2011,Jensen2011,Zhou2012,Murgas2014,Burton2015,Khalafinejad2016,Nikolov2016}. 

Ground-based detections of atomic sodium in the atmospheres of HD~189733b \citep{Redfield2008,Jensen2011} and HD~209458b \citep{Snellen2008} were enabled by the use of high-resolution spectrographs ($\lambda/\Delta\lambda \sim 50,000$). Using HARPS ($\lambda/\Delta\lambda = 115,000$) at the ESO 3.6~m telescope in La Silla (Chile), \citet{Wyttenbach2015} resolved the individual \ion{Na}{i} line profiles in the atmosphere of HD~189733b during several planetary transits. These authors established the existence of a strongly increasing temperature gradient of 0.2--0.4~K~km$^{-1}$ \citep[see also][]{Heng2015} and high-altitude winds blowing from the hot day side to the cooler night side. From a subsequent analysis of the same data set, \citet{Louden2015} measured an eastward wind blowing from the trailing limb ($-5.3^{+1.0}_{-1.4}$~\kms) to the leading limb ($+2.3^{+1.3}_{-1.5}$~\kms), after decorrelating the signal from the stellar rotation effect. A recent reanalysis of these same data \citep{Barnes2016} emphasized the role stellar activity might play in producing spurious transit signals and mimicking wind signatures in chromospherically sensitive lines, such as \ion{Ca}{ii} H and K lines \citep[e.g.,][]{RiddenHarper2016}, and in other strong features usually considered less sensitive to activity, such as \ion{Na}{i} lines. Finally, the same HARPS data allowed \citet{Cegla2016b} to pioneer a method for retrieving the stellar surface properties behind the transiting planet. This technique brings critical insights into the stellar properties during transit observations and can be used to assess how stellar rotation coupled to activity signatures (such as spots) affect transit spectroscopy.

\citet{Wyttenbach2015} established the potential of HARPS, a fiber-fed, high-resolution spectrograph stabilized in pressure and temperature, to unveil atmospheric signatures of transiting planets using only a 4-meter-class telescope. The {\it Hot Exoplanet Atmospheres Resolved with Transit Spectroscopy} survey, or HEARTS, has been undertaken to probe a large sample of gas giants in different mass and irradiation regimes using HARPS.

\begin{figure}[!htbp]
\label{fig:dearth}
\includegraphics[width=\columnwidth,trim=1.5cm 1cm 1cm 1cm]{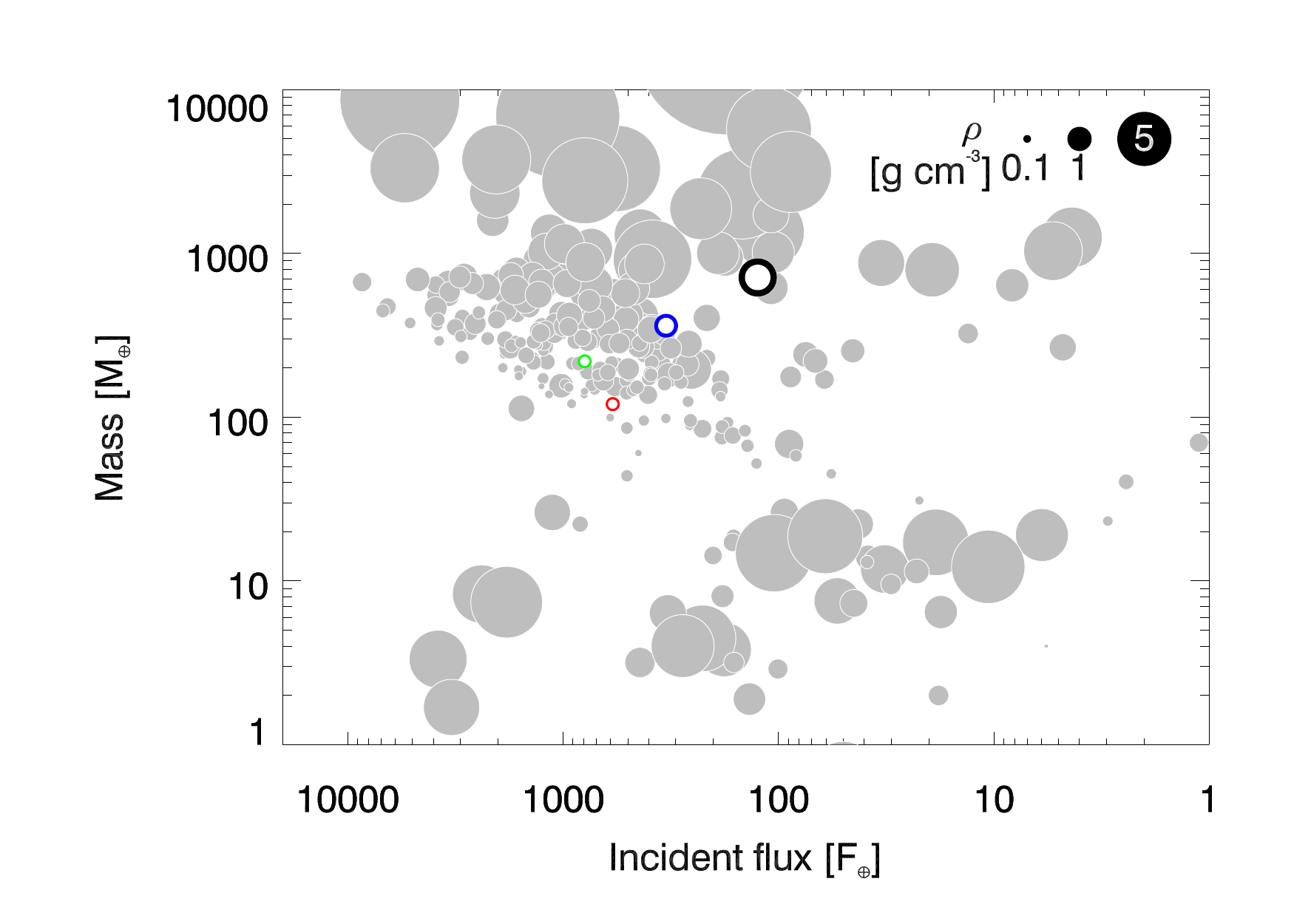}
\caption{Mass vs.\ incident flux, expressed in units of flux received at Earth, for transiting exoplanets with $V<16$ and masses determined to $<20\%$ precision (gray circles). The size of the symbols is proportional to the planet bulk density. WASP-49b is shown in red; HD~189733b, HD~209458b, and WASP-8b (see Sect.~\ref{sec:alternatives}) are in blue, green, and black, respectively.}
\end{figure}

This paper reports on the first results of the HEARTS program on exoplanet WASP-49b. WASP-49b is a gas giant that is slightly more massive than Saturn, yet larger than Jupiter \citep{Lendl2012}. The WASP-49 system properties are listed in Table~\ref{tab:w49}. This system receives 600$\times$ more flux than Earth, which is about twice the irradiation received by HD~189733b, and has an equilibrium temperature of $\sim 1,400$~K. Although the large atmospheric scale height ($\sim$700~km) of the planet makes it an a priori favorable target for transit spectroscopy, the low-resolution transmission spectrum of the planet appears flat with VLT/FORS2. This favors a cloud-dominated atmosphere \citep{Lendl2016}. Given its low mass, the planet lies at the edge of the dearth of sub-Jupiter-mass planets receiving high irradiation, which can be seen in Fig.~\ref{fig:dearth}. We present here the first high-resolution transit spectrum of the planet,  which unveils the planet upper atmospheric properties.

The paper is organized as follows. The HARPS transit observations are described in Sect.~\ref{Sec_Obs}, with details of the data reduction and telluric correction. In Sect.~\ref{Sec_CLB}, we study the stellar rotation and its potentially spurious effect on the transit spectroscopy. We present and interpret the sodium signature in the transit spectrum of the planet in Sect.~\ref{Sec_TransSpec}.

\section{Observations and data reduction}\label{Sec_Obs}

\subsection{HARPS observations}
We observed the G6V star WASP-49 during three transits of its hot gas giant planet, WASP-49b \citep{Lendl2012}, with the HARPS \citep[High-Accuracy Radial-velocity Planet Searcher;][]{Mayor2003} spectrograph mounted on the ESO 3.6 m telescope in La Silla (Chile). These observations have been performed during the first semester of our HEARTS survey (ESO program 096.C-0331; PI: Ehrenreich), an atmospheric survey of hot exoplanets started in the wake of the spectrally resolved detection of sodium in the atmosphere of HD~189733b with HARPS \citep{Wyttenbach2015}. 

Our observational strategy is to record, for each planet, several transits to assess the reproducibility of the detected spectroscopic signatures. We dedicate a whole night of continuous observations for each transit to retrieve a good out-of-transit baseline and build an accurate out-of-transit master spectrum. A high signal-to-noise ``master out'' spectrum is key to the differential technique first presented in \citet{Wyttenbach2015} and used throughout this work. A good baseline also allows for an efficient correction of telluric lines.

We observed transits of WASP-49b on 6 December 2015, 31 December 2015 and 14 January 2016 (see the log of observations in Table~\ref{tab:log}), where one fiber is on the target (fiber A) and the other monitors the sky (fiber B). In total, we recorded 126 spectra of 600~s exposure time. Our analysis is based on 125 spectra; we discarded one spectrum that was polluted by twilight. The typical signal-to-noise ratio (S/N) calculated in the continuum around 590 nm are between 20 and 40. During the second night, a series of eight spectra have lower S/N values in comparison to other spectra (with values between 10 and 20). This is probably due to cloud passages. Our results on the stellar rotation analysis (Sect.~\ref{Sec_CLB}) are not affected by taking into account or discarding these spectra. We also computed the transmission spectrum with and without taking these spectra into account. For the transmission spectrum analysis, we decided to discard the eight low S/N spectra. Indeed for low S/N, the line cores of strong stellar absorption lines (as \ion{Na}{i}) have almost no flux recorded, thus the line cores are dominated by readout noise and background residuals, which reflect in a systematic error dominating the transmission spectrum. Therefore, for the second night we removed the low S/N series of eight spectra, which corresponds to three spectra in transit and five spectra out of transit.

The transit coverage can be appreciated in the lower panel of Fig.~\ref{DACEsol}. Our observations cover three full planetary transits with baselines of $\sim$1--3~h before and $\sim$2--3~h after each transit. We identified 39 spectra partially or fully in transit (i.e., with a part of the exposure between the first and fourth contacts). The 86 remaining spectra are considered out of transit.

\subsection{Data reduction}
The HARPS observations were automatically reduced with the last version of the HARPS Data Reduction Software (DRS version 3.5). The DRS performs an order-by-order extraction of the spectra, which are then flat-fielded with the daily calibration set. A blaze correction is applied to each spectral order together with a wavelength calibration. The wavelengths are given in the air and the solar system barycenter rest frame is chosen as reference. The 72 spectral orders of each two-dimensional echelle spectrum are merged and resampled into a one-dimensional spectrum (this process ensures flux conservation). The final reduced HARPS spectra sample the region between 380~nm and 690~nm with a 0.01~$\AA$ wavelength step. The spectral resolution is $\lambda/\Delta\lambda = 115,000$ or 2.7~\kms. Cross-correlation functions (CCFs) are constructed by correlating the spectra with a G2-type stellar mask \citep{Pepe2002}. The radial velocities of the star are computed with a Gaussian fit to the CCFs. In addition, the absolute calcium activity index ($\log R'_{HK}$) is computed for each spectrum by the DRS following \citet{Noyes1984}. We computed an average $\log R'_{HK}$ index from all individual values after verifying that the measurements do not show any significant systematic trend with the S/N.

\begin{table}
\caption{Adopted values for the orbital and physical parameters of the WASP-49 system.}
\begin{center}
\begin{tabular}{lc}
\hline
Parameter & Value \\
\hline
\leavevmode\\
\multicolumn{2}{l}{Star: WASP-49}\leavevmode\\
$V$ mag                 & 11.36\\
Spectral type           & G6\,{\sc v}\tablefootmark{\textasteriskcentered}\\
$M_{\star}$             & $1.003\pm0.10$~\Msun\tablefootmark{\textasteriskcentered}\\
$R_{\star}$             & $1.038\pm0.038$~\Rsun\tablefootmark{\textasteriskcentered}\\
$T_\mathrm{eff}$        & $5600\pm150$~K\tablefootmark{\textasteriskcentered}\tablefootmark{\dag}\\
$\log g$                & $4.5\pm0.1$\tablefootmark{\textasteriskcentered}\tablefootmark{\dag}\\
$\log R'_{HK}$          & $-5.17\pm0.02$\tablefootmark{\ddag}\\ 
\leavevmode\\
\multicolumn{2}{l}{Transit}\leavevmode\\
$T_0$                   & $2456267.68389\pm0.00013~\mathrm{BJD_{tdb}}$\tablefootmark{\textasteriskcentered}\\
$P$                     & $2.7817362\pm0.0000014$~d\tablefootmark{\textasteriskcentered}\\
$(R_p/R_\star)^2$       & $0.01345\pm0.00017$\tablefootmark{\textasteriskcentered}\\
$\Delta T_\mathrm{1-4}$ & $2.14\pm0.01$~h\tablefootmark{\textasteriskcentered}\\
$b$                     & $0.7704^{+0.0072}_{-0.0077}~R_{\star}$\tablefootmark{\textasteriskcentered}\\
$a$                     & $0.03873\pm0.00130$~au\tablefootmark{\textasteriskcentered}\\
$i$                     & $84.48\degr\pm0.13\degr$\tablefootmark{\textasteriskcentered}\\
$u_1,u_2$               & $0.34, 0.28$\tablefootmark{\textasteriskcentered}\tablefootmark{\S}\\
\\
\multicolumn{2}{l}{Radial velocity}\\
$K_1$                   & $57.5\pm2.1$~m~s$^{-1}$\tablefootmark{\ddag}\\
$e$                     & 0\tablefootmark{\ddag}\\
$\omega$                & $90\degr$\tablefootmark{\ddag}\\
$\gamma$                & $41\,726.1\pm1.1$~m~s$^{-1}$\tablefootmark{\ddag}\\
\leavevmode\\
\multicolumn{2}{l}{Planet: WASP-49b}\\
$M_p$                   & $0.399\pm0.030$~\Mjup\tablefootmark{\ddag}\\
$R_p$                   & $1.198\pm0.047$~\Rjup\tablefootmark{\textasteriskcentered}\\
$\rho$                  & $0.288\pm0.006$~g~cm$^{-3}$\tablefootmark{\ddag}\\
$g_\mathrm{surf}$       & $689\pm11$~cm~s$^{-2}$\tablefootmark{\ddag}\\
$T_\mathrm{eq}$         & $1400\pm80$~K\tablefootmark{\ddag}\tablefootmark{\P}\\ 
$H$                     & $730\pm40$~km\tablefootmark{\ddag}\tablefootmark{\P}\\
$2R_pH/R_\star^2$       & $230\pm17$~ppm\tablefootmark{\ddag}\\
\hline
\end{tabular}
\end{center}
\tablefoot{}
\tablefoottext{\textasteriskcentered}{\citet{Lendl2016}.}
\tablefoottext{\dag}{\citet{Stassun2016}.}
\tablefoottext{\ddag}{This work.}
\tablefoottext{\S}{Quadratic limb darkening from the NGTS filter.}
\tablefoottext{\P}{For the planetary equilibrium temperature $T_\mathrm{eq}$ and the atmospheric scale height $H$, we assume an albedo $A=0$, a redistribution factor $f=1$ and a mean molecular weight $\mu_{\mathrm{atm}}=2.3$.}
\label{tab:w49}
\end{table}

\begin{table*}
\caption{Log of observations.}
\begin{center}
\begin{tabular}{lccccccc}
\hline
\rule[0mm]{0mm}{5mm}    & Date & $\#$ Spectra$^1$ & Exp.\ Time [s] & Airmass$^2$ & Seeing & $S/N$         & $S/N$\\
                        &      &                  &                &             &        & continuum$^3$ & Na core$^4$\\
\hline
$\mathrm{Night\ 1}$ & 2015-12-06 & 41 (14/27) & 600 & 1.9--1.02--1.3 & not recorded & 20.7--32.6 & 5--7\\
$\mathrm{Night\ 2}$ & 2015-12-31 & 44 (12/32) & 600 & 1.4--1.02--2.2 & 0.5--1.3 & 10.2--38.1 & 2--8\\
$\mathrm{Night\ 3}$ & 2016-01-14 & 40 (13/27) & 600 & 1.2--1.02--2.3 & 0.5--1.0 & 21.8--39.5 & 5--9\\
\hline
\end{tabular}
\tablefoot{$^1$In parenthesis: the number of spectra taken during the transit and outside the transit, respectively. $^2$The three figures indicate the airmass at the start, middle and end of observations, respectively. $^3$The signal-to-noise ($S/N$) ratio per pixel extracted in the continuum near 590~nm. $^4$The $S/N$ ratio in the line cores of the \ion{Na}{i} D doublet.}
\end{center}
\label{tab:log}
\end{table*}

Since WASP-49 is a relatively faint target, we paid particular attention to systematic effects potentially occurring at low flux levels. We investigated whether contamination by moonlight could have added a spurious background on top of the WASP-49 spectrum. To do so we used the data from fiber B, which was monitoring the sky background simultaneously to the WASP-49 observations. A cross-correlation of the sky spectrum with a G2 template yielded no detectable signal during any of the observing nights. We can therefore exclude any moonlight pollution effects. We also examined whether the automatic background subtraction performed by the HARPS pipeline could have systematically over- or underestimated the diffuse background and dark current level on the detector. By taking the average flux level on fiber B during the different nights, we found that background residuals are at the level of 0.6, 0.8 and 1 electron per extracted pixel with a standard deviation over a night of 0.2--0.3 electron (for both the in- and out-of-transit spectra). We simulated the impact of the background residuals on the transmission spectrum in the \ion{Na}{i} D line cores. To do so, we measured the actual in-transit and out-of-transit fluxes (on the fiber A) in the continuum around the \ion{Na}{i} D lines and compared these with the line core fluxes by assuming the \ion{Na}{i} D lines contrast to be 95\%. We artificially added the offset measured on the fiber B to each flux. We found that the error on the background subtraction causes an artificial absorption of about 0.3--0.5\% in the transmission spectrum. This effect is buried within the noise of our final transmission spectrum. We conclude that the background correction uncertainty does not significantly affect the WASP-49 spectra and transmission spectrum.

\subsection{Telluric correction}\label{subSec_tellcorr}
Ground-based observations are affected by the atmosphere of the Earth. The atmospheric constituents contaminate the HARPS spectra with a diversity of telluric absorption lines. Telluric water and molecular oxygen imprint a time-variable contamination to the optical stellar spectra, with line contrasts between $\sim$100\% (full absorption, e.g.\ the $\mathrm{O_2}$ B band) and $\la$1\% (the so-called microtellurics). The atmospheric absorption of the Earth changes during a night because of airmass and water column variations. This telluric contamination can be mitigated by an appropriate subtraction technique, such as that described by \citet{Snellen2008}, \citet{Astudillo-Defru2013}, or \citet{Wyttenbach2015}. The advantage of these techniques is that they are independent of Earth atmospheric models and they do not require observations of a
telluric standard star. The telluric absorption signatures can be reduced by a factor $\sim$10 to 50, depending on the line contrasts and typical photon noise level. In the case of WASP-49, the individual spectrum S/N ratios (the measured S/N ratios lie between 20 and 40) cause a typical standard deviation around the continuum between 3 and 5\%, preventing us from verifying the quality of the telluric correction below that level. 
Another aspect is that the telluric lines move with respect to the stellar spectrum according to the barycentric Earth radial velocity (BERV). During the observations, the BERV had values of $6.6$, $-3.5$, and $-9$~\kms. This means that the telluric lines shift by $0.3~\AA$ relative to the stellar spectrum between the first and third night. Thus, we verified that, in all nights, the sodium D$_2$ line core is free of telluric contamination. On the other hand, the D$_1$ line core is affected by some telluric water transitions, but we did not find obvious residual features in the transmission spectrum. Despite these aspects, we were able to show that the telluric absorption in the 5875 to 6000~$\AA$ region linearly follows the airmass during each night, meaning no significant water column variation was detected. This allowed us to follow the same telluric correction method as in \citet{Wyttenbach2015}. We combine the out-of-transit spectra to build a telluric spectrum for each night and use it to correct all individual spectra. This means that all spectra, as well as the final transmission spectra, are corrected from the telluric lines down to the noise level.

To ensure that our transmission spectrum is not contaminated by telluric sodium (of extraterrestrial origin), we visually inspected each spectrum in the \ion{Na}{i} doublet region for both fiber A and B. We did find traces of telluric sodium in a half-dozen spectra at a level of 20 to 30 electrons per extracted pixel, uncorrelated with airmass. All the other spectra had no significant telluric sodium flux above the 5 to 10 electrons level. The randomness of the telluric sodium emission makes the transmission spectra difficult to correct around this very specific spectral region. By looking at the master spectra of fiber B (sky only), we estimated that the impact of telluric sodium on the transmission spectrum could be up to 0.3--0.5\% in absorption. However, this absorption feature is shifted by 0.8~\AA\ (40~\kms) from the stellar line cores due to the high systemic velocity of the star. We can therefore safely exclude telluric sodium as a source of systematic error in our observations.

\begin{figure}[tbp]
\centering
\includegraphics[width=0.47\textwidth]{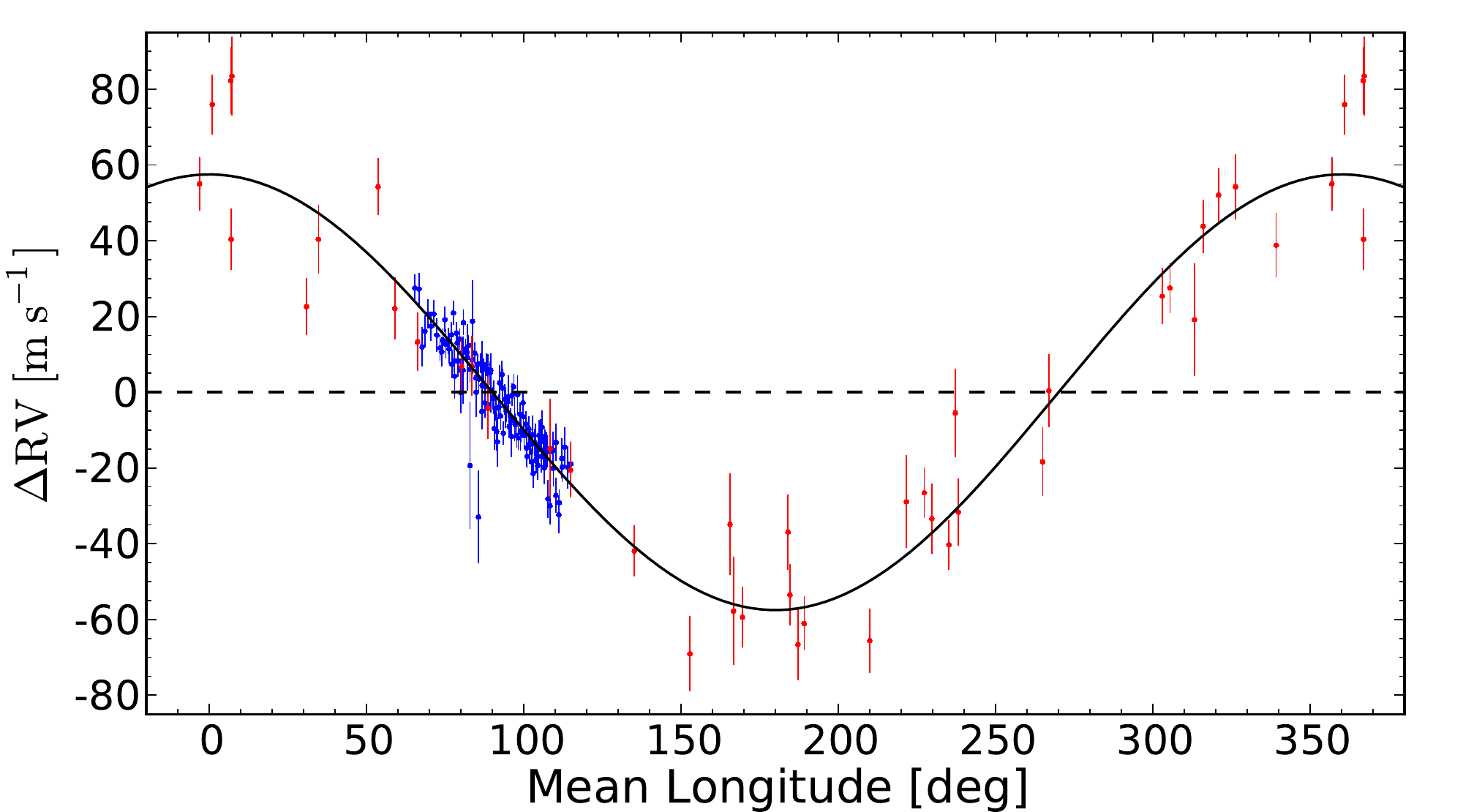}
\includegraphics[width=0.47\textwidth]{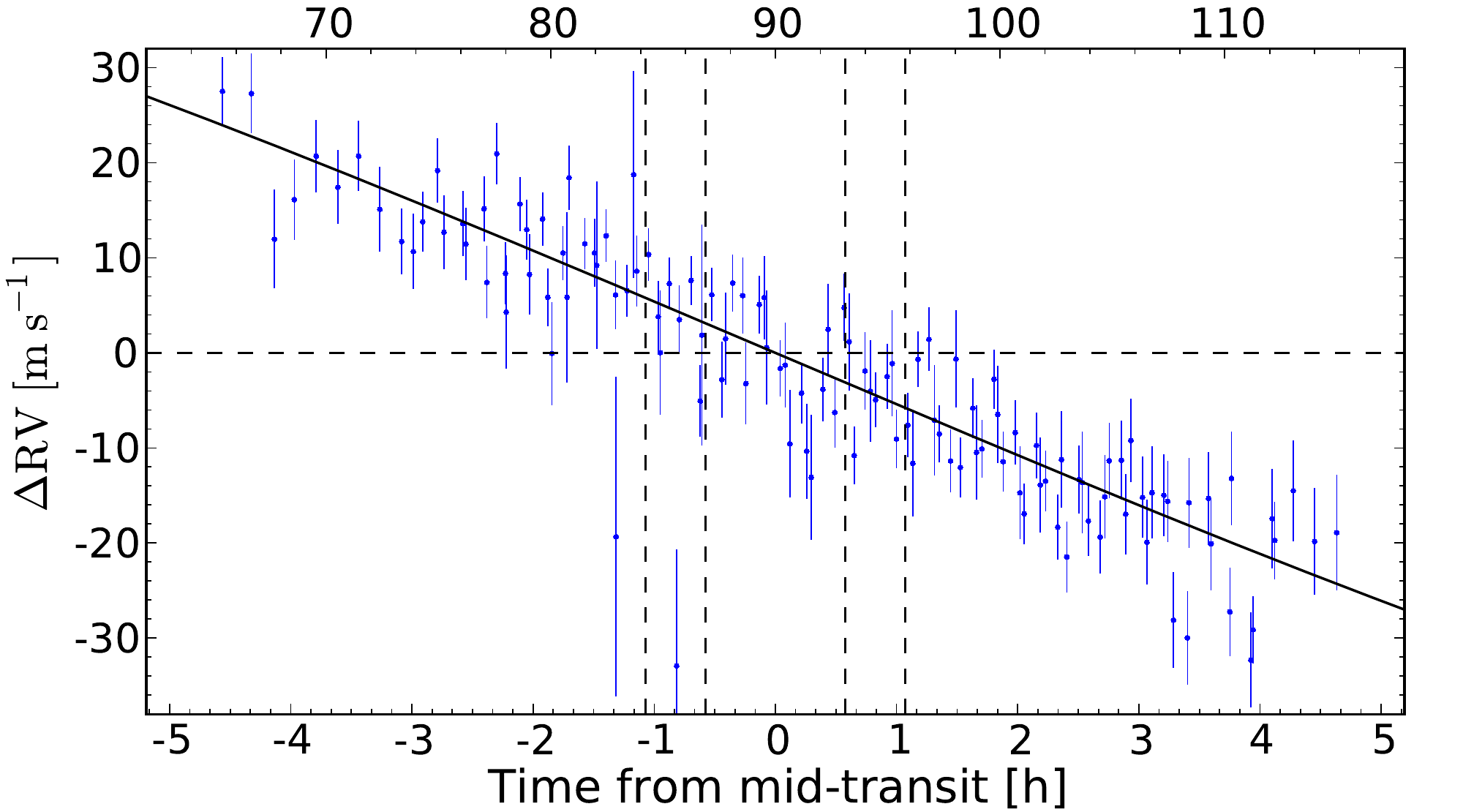}
\caption{Upper panel: Radial velocity measurements of WASP-49 with CORALIE (red points) and HARPS (blue points). The measures are fitted with a Keplerian orbit onto the DACE platform (see Sect.~\ref{Dace_sol}). The result is compatible with the circular orbit shown here. Lower panel: Zoom in on the three HARPS sequences only. The transit timing is indicated with dashed vertical lines representing the four contact points. No obvious Rossiter-McLaughlin effect is noticeable.}
\label{DACEsol}
\end{figure}

\subsection{Orbit refinement with DACE}\label{Dace_sol}
We use the DACE platform\footnote{The Data \& Analysis Center for Exoplanets (DACE) platform is available at https://dace.unige.ch} to analyze the archival CORALIE \citep{Lendl2012} radial velocity data and our new HARPS data of WASP-49. An orbit is first fitted starting from an analytical determination of the orbital elements using Fourier analysis \citep{Delisle2016}. The orbital parameters of one Keplerian orbit are computed from the Lomb-Scargle periodogram \citep{Zechmeister2009} of the radial velocity data by selecting and extracting the periods and amplitudes of the strongest peak and its first harmonic. Then, the Keplerian solution is adjusted to the data with a Levenberg-Marquardt algorithm. Starting with the latter solution, we use DACE to perform a Markov chain Monte Carlo (MCMC) algorithm on the radial velocity data \citep{Diaz2014,Diaz2016}. For the MCMC inputs, we took Gaussian priors for the mid-transit time and the orbital period with values and 1$\sigma$ uncertainties coming from the high-precision photometry of \citet{Lendl2016}. All the other priors (on the semi-amplitude, eccentricity, and argument of periapsis, and the offset between the two instruments) are uniform. Finally, to model correctly the errors on the radial velocities, a quadratically additive error is applied to each data point. A unique value is taken for each instrument; for the MCMC we put a uniform prior on the two instrumental error parameters. The results are shown in Table~\ref{tab:w49} and in Appendix~\ref{app:MCMC_DACE}. The posterior distributions strongly favor a null eccentricity. Thus, we fix the eccentricity to zero for the subsequent analysis (see Fig.~\ref{DACEsol} and Table~\ref{tab:w49}).

\begin{figure*}[t!]
\centering
\includegraphics[width=0.97\textwidth]{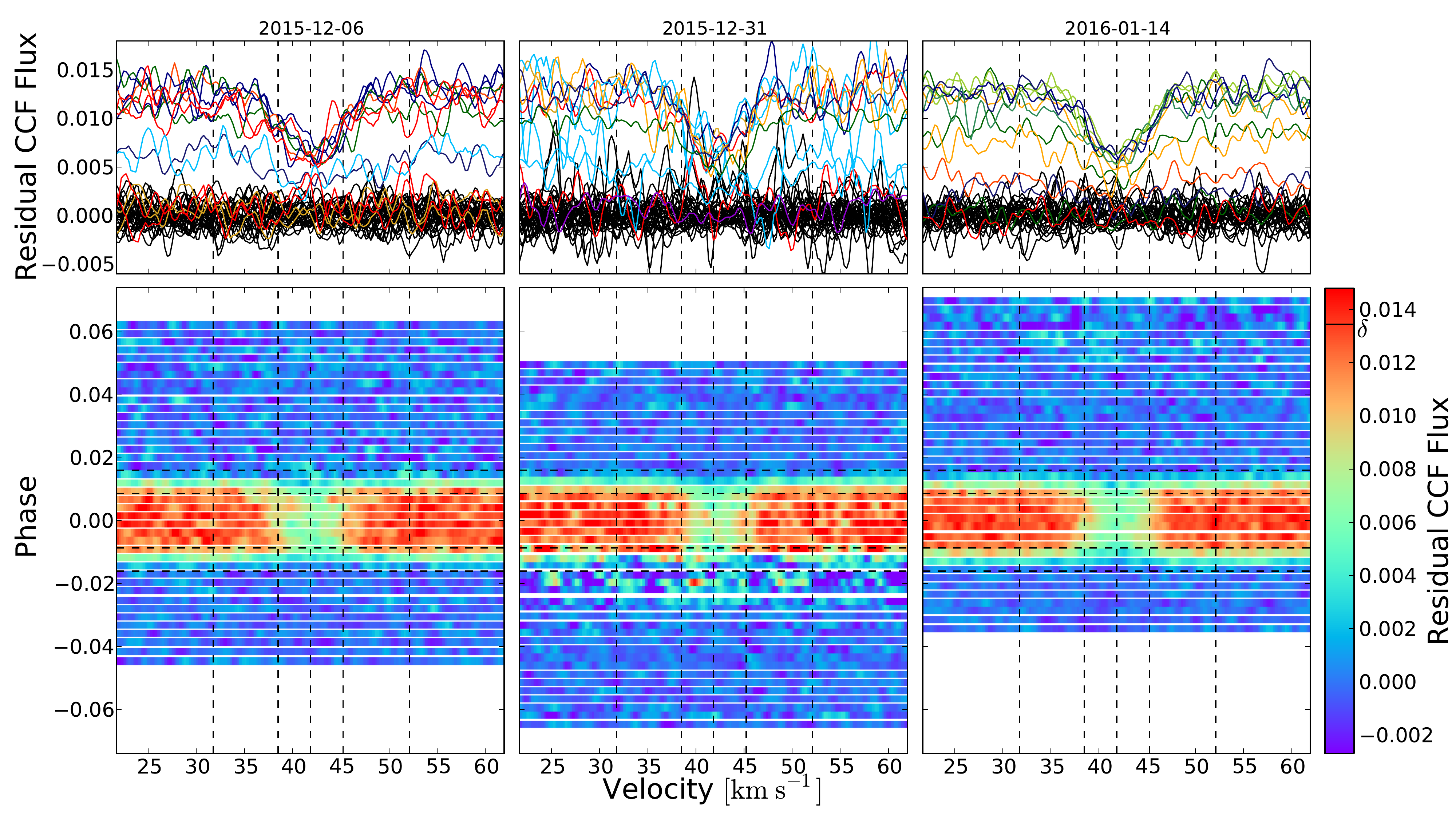}
\caption{CCFs of the stellar surface occulted by the planet, $\mathrm{CCF}_{\mathrm{res}}$, for our three nights of observations (with the date indicated as column title from left to right). Top: Residual CCFs are shown in black for the out-of-transit data and randomly colored for the in-transit data. The vertical dashed lines show the normalization regions (indicated by the two outer lines and defined by $v\leq31.76$~\kms or $v\geq52.13$~\kms), the FWHM of the master CCF profile (middle two lines at $v=\gamma\pm3.37$~\kms), and the systemic velocity $\gamma$ (central line at $v=41.73$~\kms). Bottom: Residual map of the time series $\mathrm{CCF}_{\mathrm{res}}$ in phase and color-coded by residual flux. The horizontal dashed lines represent the four contact points of the transit. As for the radial velocity series (Fig.~\ref{DACEsol}), no obvious traveling planet signature is noticeable (see Fig.~\ref{ccf_vel}).}
\label{ccf_resid}
\end{figure*}

\section{``Reloaded'' Rossiter-McLaughlin analysis}\label{Sec_CLB}

In a transit event, the Rossiter-McLaughlin (RM) effect is a manifestation of stellar rotation. It imprints a characteristic signal in the radial velocity curve, whose amplitude and shape depend on the stellar rotational velocity and the sky-projected angle between the stellar spin axis and the normal to the orbital plane, respectively. The RM effect can create a residual signal that is visible in the transmission spectra obtained with high-resolution spectrographs \citep{Louden2015,Brogi2016}. Conversely, our high-resolution spectra can be used to infer the underlying properties of the stellar surface and constrain differential rotation \citep{Cegla2016b}, making it possible to measure its impact on the transmission spectra.

Here, we follow the methodology of \citet{Cegla2016b}. This technique, which uses the planet as a probe to resolve the stellar surface along the transit chord, is summarised in Sect.~\ref{paap}. This methodology measures the stellar projected rotation (and any other velocity fields emerging from the stellar surface) along the chord of the planetary transit \citep[][see also \citet{Czesla2015}]{Cegla2016b,Bourrier2016}. This ``reloaded RM'' approach, in contrast to the classical RM analysis \citep{Queloz2000,Triaud2010,Albrecht2012,Brown2016}, provides us with
\begin{enumerate}
\item A direct measurement of the projected velocity field at the surface of the star (see Sect.~\ref{subsubSec_CLB_MCMC})
\item The local spectrum (and CCF) of the region of the stellar disk hidden by the planet during the transit (see Sect.~\ref{subsubSec_CLB_FWHM}).
\end{enumerate}
This information then allows us to estimate the impact of the RM effect on the transmission spectrum

\subsection{Planet as a probe: resolving the stellar surface}\label{paap}
We present here the data treatment to retrieve the stellar surface properties, notably the surface radial velocity $v_{\mathrm{surf}}$, following the methodology of \citet{Cegla2016b}.

We begin by removing the orbital radial velocities (corresponding to the circular orbit retrieved in Sect.~\ref{Dace_sol}) from each observed cross-correlation function (hereafter, $\mathrm{CCF}(t)$). Then, the $\mathrm{CCF}(t)$ were all continuum-normalized, where the continuum region was defined in the velocity range $v\leq31.76$~\kms\ or $v\geq52.13$~\kms. This yields the normalized $\widetilde{\mathrm{CCF}}(t)$. A master out-of-transit CCF, designated as $\widetilde{\mathrm{CCF}}_\mathrm{out}$, is built by summing and then continuum normalizing all the out-of-transit $\mathrm{CCF}(t)$. We finally compute the residual CCFs as follows:

\begin{equation}
\mathrm{CCF}_{\mathrm{res}}(t) = \widetilde{\mathrm{CCF}}_\mathrm{out} - [1 - \delta(t)] \cdot \widetilde{\mathrm{CCF}}(t),
\end{equation}

where $1-\delta(t)$ represents the value of the photometric light curve, and $\delta(t)$ is the transit depth at time $t$. The light curve is computed with a \citet{Mandel2002} model using the \texttt{batman} code \citep{Kreidberg2015} and the parameters from Table~\ref{tab:w49}. Note that although the limb-darkening is not a dominant effect \citep{Cegla2016b}, we choose the limb-darkening parameters measured by \citet{Lendl2016} in the NGTS filter, because the latter has the broadest wavelength coverage in common with HARPS. The residual CCFs are shown in Fig.~\ref{ccf_resid}.

A Gaussian fit to the residual CCFs using a Levenberg-Marquardt algorithm allows us to measure the centroid, contrast, and full width at half maximum (FWHM), together with error bars, of each individual $\mathrm{CCF}_{\mathrm{res}}$. Only one in three CCF data points is used in the fit due to the oversampling of the CCFs produced by the HARPS pipeline. The centroid value represents the radial velocity of the local stellar surface $v_{\mathrm{surf}}$. This is meaningful only for the in-transit residual CCFs, where the $\mathrm{CCF}_{\mathrm{res}}$ represent the local line profile from the planet-occulted regions. We also measure the dispersion in the continuum of the residual CCFs. A given $\mathrm{CCF}_{\mathrm{res}}$, and thus its $v_{\mathrm{surf}}$, is considered detected if its contrast is at least three times larger than the dispersion in the continuum (see Fig.~\ref{ccf_vel}). Indeed, below this threshold, the Gaussian fit has a high probability to converge onto a noise feature, which does not correspond to a physical signal. In practice, this led us to discard observations with a limb angle $\mu=\cos\theta$ lower than $\sim 0.25$--$0.3$, and to keep 25 $v_{\mathrm{surf}}$ measurements for our analysis of the spin-orbit alignment.

\subsubsection{Analysis of the local stellar surface velocities}\label{subsubSec_CLB_MCMC}

\begin{figure}[tbp]
\centering
\includegraphics[width=0.47\textwidth]{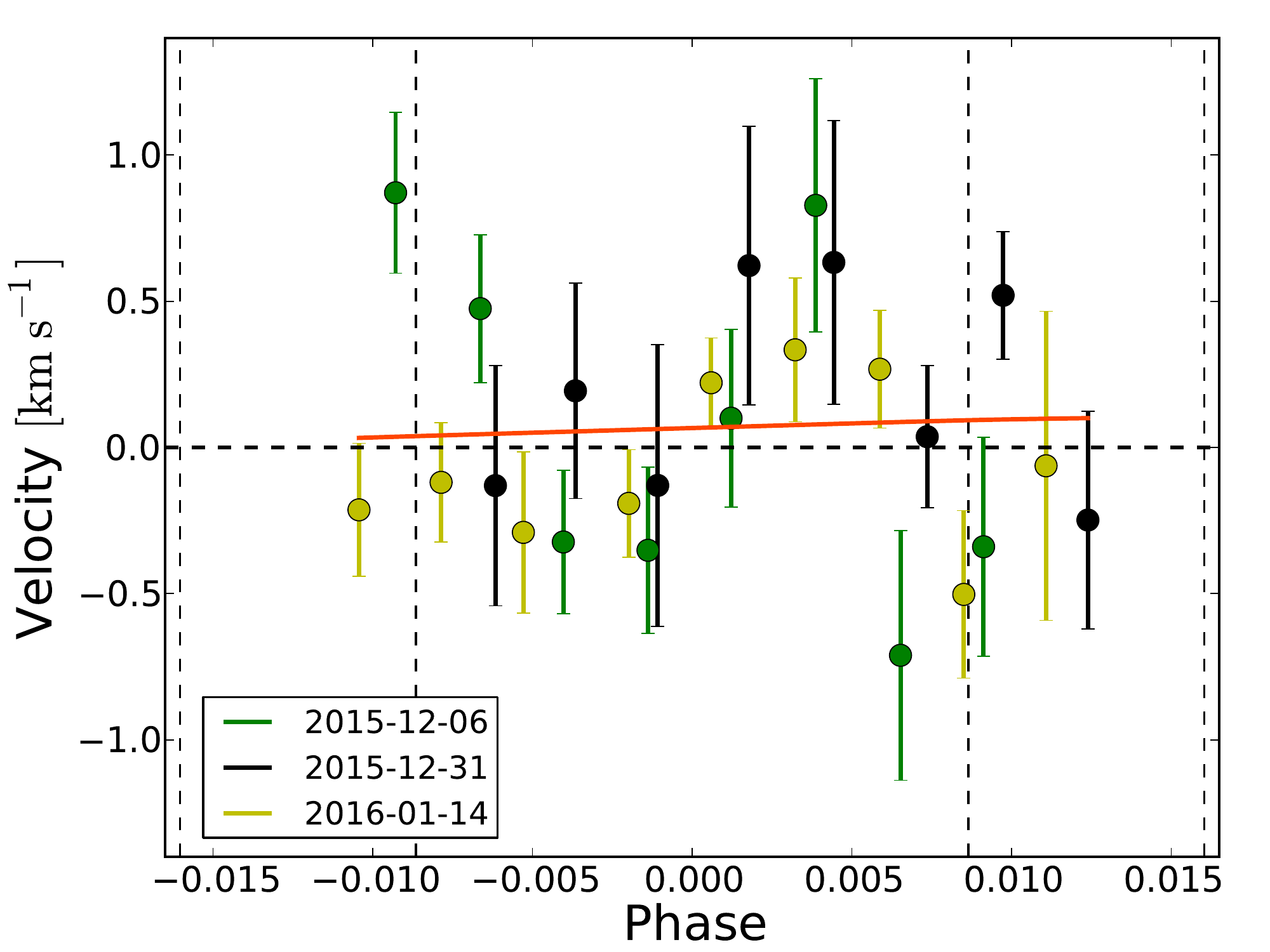}
\caption{Radial velocity of the stellar surface occulted by the planet for our three nights of observations. The vertical dashed lines represent the four contact points of the transit. The best solid-body rotation model is shown in orange. However, the $v_{\mathrm{surf}}$ are better modeled by a null stellar surface radial velocity (horizontal dashed line), i.e., with a nondetected Rossiter-McLaughlin effect. There is a presence of a correlated pattern, near the transit center and common to all nights, which remains unexplained.}
\label{ccf_vel}
\end{figure}

The measured local stellar surface velocities $v_{\mathrm{surf}}$ are presented in Fig.~\ref{ccf_vel}. They can be directly interpreted using the \citet{Cegla2016b} model, which computes the brightness-weighted average radial velocity behind the planet (see their Section 2.2 for more details). The radial velocities of the regions occulted by the planet along the transit chord depend on the transit geometry and the stellar and planetary physical parameters. Some of these parameters (in particular the limb-darkening coefficients) are fixed to values listed in Table~\ref{tab:w49}, while others are determined by the fit to the measured radial velocities: the sky-projected spin-orbit angle $\lambda$, the angle between the line of sight and the stellar spin axis $i_{*}$, the stellar equatorial rotational velocity $v_\mathrm{eq}$, a factor $\alpha$ describing a solar-like differential rotation law \citep{Schroeter1985}, and a contribution from the convective velocities at the stellar surface. Under the assumption of solid-body rotation, $\alpha$ is set to 0 and there is a degeneracy between $v_\mathrm{eq}$ and $i_{*}$ (they are thus replaced by $v \sin i_{*}$ in the fitted parameters).

In each of the scenarios considered hereafter, we fitted for the free parameters using a Metropolis-Hastings MCMC algorithm. We ran several chains started from different random positions in the parameter space, and adjusted the jump size to get an acceptance rate of $\sim25\%$. To better sample the posterior distribution in the case of nonlinear correlations, we applied an adaptive principal component analysis to the chains \citep{Bourrier2015,Cegla2016b}. We checked that all chains converged to the same solution, before thinning them using the maximum correlation length of all parameters. Finally, we merged the thinned chains so that the posterior probability distributions contained a sufficient number of independent samples. The best-fit values for the model parameters were set to the medians of the posterior distributions, and their 1$\sigma$ uncertainties are evaluated by taking limits at 34.15\% on either side of the median.

\begin{figure}[tbp]
\centering
\includegraphics[width=0.47\textwidth]{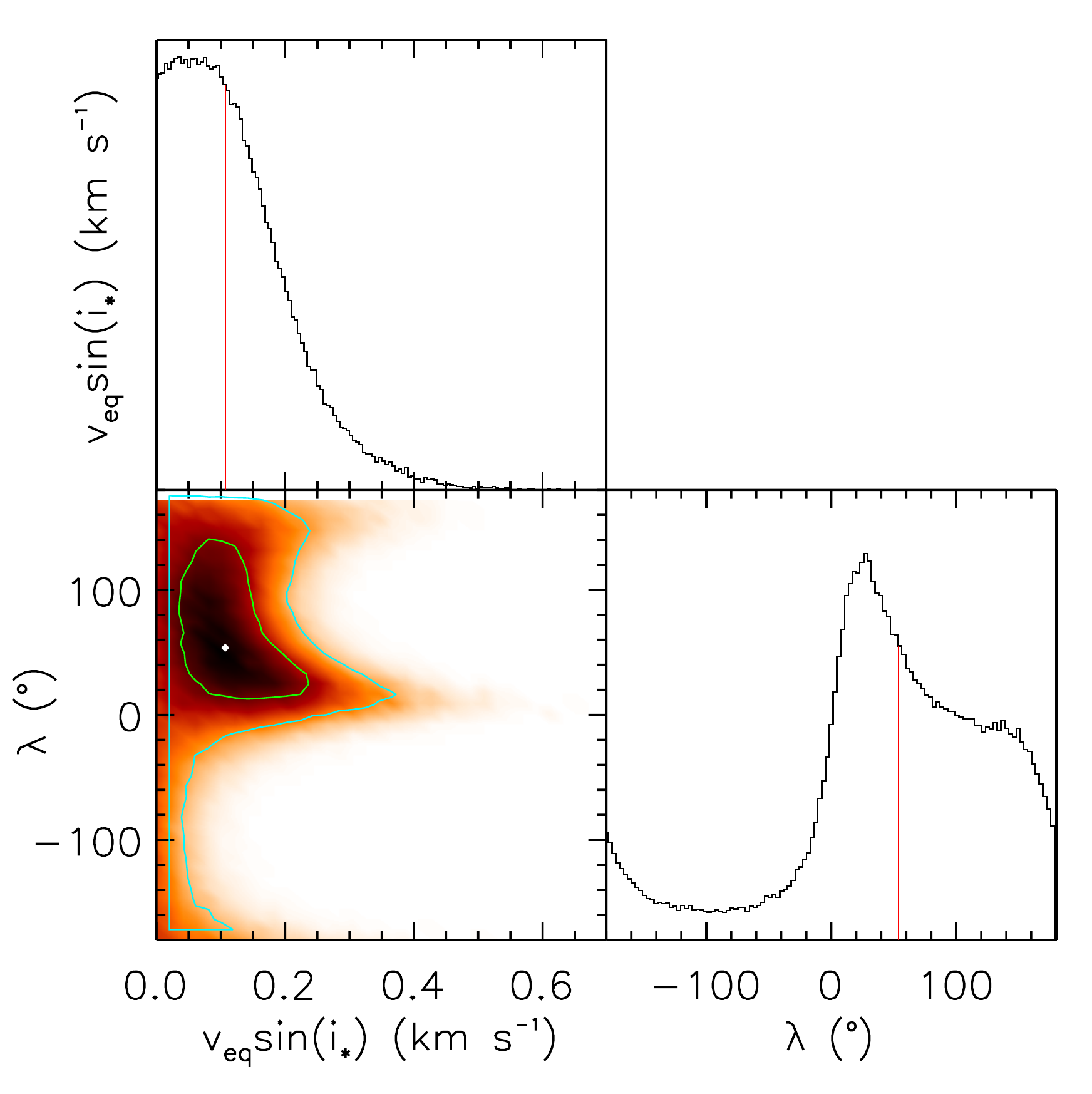}
\caption{Correlation diagrams for the posterior probability distributions of the solid-body stellar rotation model parameters. Green lines show the 1 and 2~$\sigma$ simultaneous 2D confidence regions that contain 39.3\% and 86.5\%, respectively, of the accepted steps. 1D histograms correspond to the distributions projected on the space of each parameter. The red lines and white point show median values.}
\label{ccf_mcmc}
\end{figure}

We start by fitting a solid-body rotation model to the data. The correlation diagram for the MCMC-derived parameters are shown in Fig.~\ref{ccf_mcmc}. The best-fit corresponds to $v \sin i_{*} = 0.11^{+0.10}_{-0.07}$~\kms\ and $\lambda = 54^{\degr+79\degr}_{-58\degr}$, and yields $\chi^2=44.7$ for $\nu=25$ degrees of freedom (reduced $\chi^2_\nu=1.8$). The value of the Bayesian information criterion (BIC) for this model is 51.2. The uncertainties on the radial velocity measurements are too large to constrain any convective velocity contribution from the stellar surface. We then tried to fit a more complex model that includes differential rotation, but found that the data does not allow us to meaningfully constrain the $\alpha$ parameter, which exhibits a roughly flat posterior distribution. Finally, we compared the solid-body model to the null hypothesis, i.e. that the stellar surface velocity field along the transit chord cannot be significantly detected. The null velocity model yields $\mathrm{BIC}=\chi^2=46.4$ ($\chi^2_\nu=1.9$). This is lower than the BIC for the solid-body model, showing that including stellar rotation does not bring significant improvement to the fit. We thus conclude that the surface velocity field of WASP-49 is not significantly detected. This can have two interpretations. First, the star is seen pole-on (i.e., the stellar spin axis is aligned with the line of sight), in which case the projected rotational velocities would be close to zero even if the star was a fast rotator. Second, the star is a very slow rotator. In both cases the MCMC results can be used to put a 3$\sigma$ upper limit on $v \sin i_{*} \leq 0.46$~\kms. A correlated pattern is present near the transit center in all nights (Fig.~\ref{ccf_vel}) and could arise from a persistent surface feature on one of the stellar poles. We do not explore this hypothesis further here, lacking an appropriate physical model for such a feature and its radial velocity signature.

\subsubsection{Analysis of the local stellar surface line profiles}\label{subsubSec_CLB_FWHM}
\begin{figure}[tbp]
\centering
\includegraphics[width=0.47\textwidth]{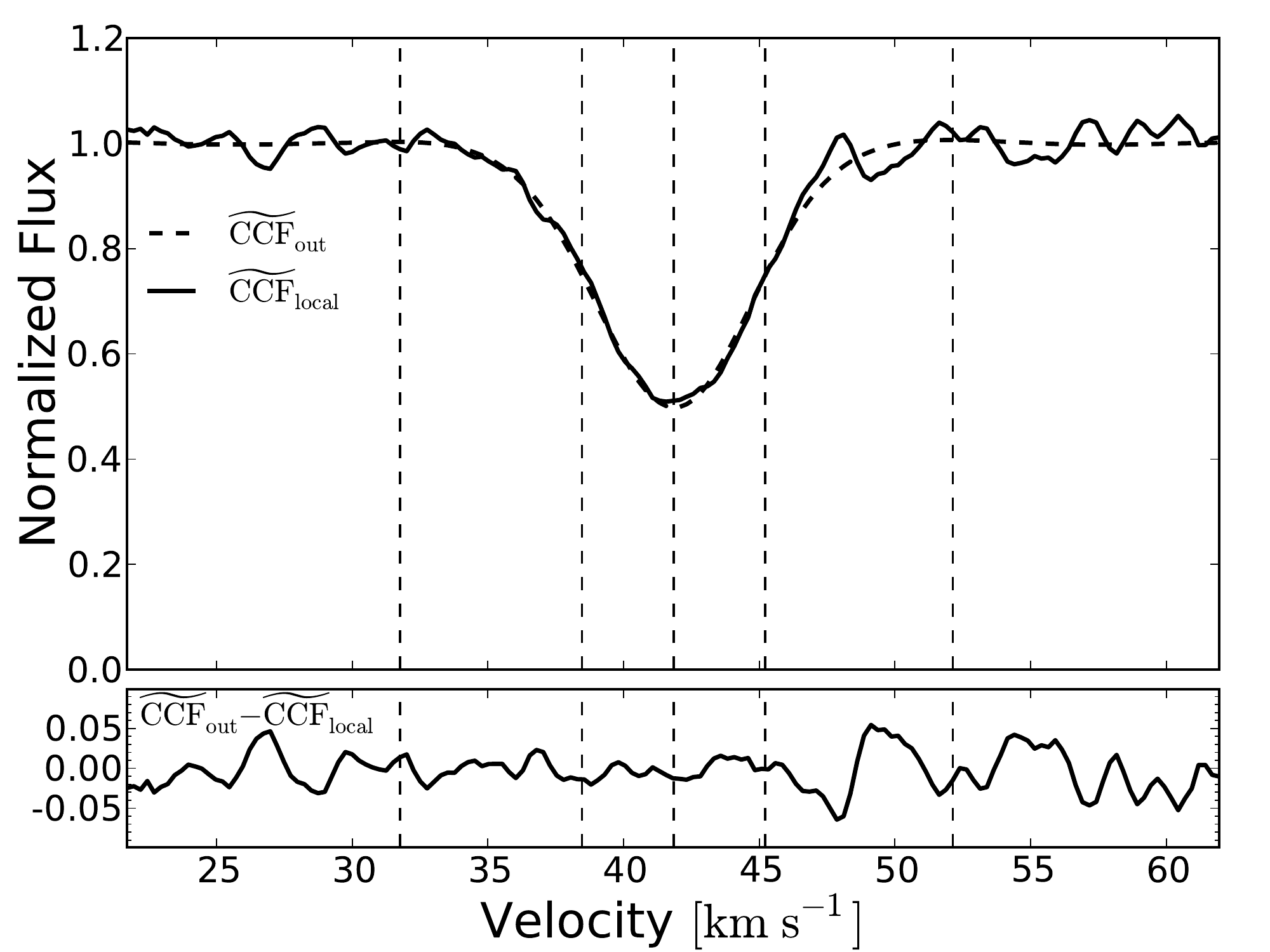}
\caption{Upper panel: Comparison between the disk-integrated master $\widetilde{\mathrm{CCF}}_\mathrm{out}$ (dashed line) and the local master $\widetilde{\mathrm{CCF}}_{\mathrm{local}}$ (full line). Lower panel: Difference between both CCFs. The vertical dashed lines show the normalization regions (outer two lines at $v=31.76$~\kms and $v=52.13$~\kms), FWHM of the $\widetilde{\mathrm{CCF}}_\mathrm{out}$ (middle two lines at $v=\gamma\pm3.37$~\kms), and systemic velocity $\gamma$ (central line at $v=41.73$~\kms).}
\label{ccf_master}
\end{figure}

With the residual CCF method from \citet{Cegla2016b}, we are able to probe the stellar surface along the transit chord. For example, one can compare the local CCF contrast and FWHM to 3D magnetohydrodynamic (MHD) simulations and seek local CCF shape variations across the stellar surface \citep[see][]{Bourrier2016}. Here, we extend this method further by computing a master local CCF, designated  $\widetilde{\mathrm{CCF}}_{\mathrm{local}}$, to increase the S/N and compare its properties to those of the disk-integrated CCF given by $\widetilde{\mathrm{CCF}}_\mathrm{out}$. In doing so we obtain an average local CCF profile that neglects potential variations across the stellar surface. The comparison of the two master CCFs yields an independent measurement of $v \sin i_{*}$. The $\widetilde{\mathrm{CCF}}_{\mathrm{local}}$ is obtained by shifting each in-transit $\mathrm{CCF}_{\mathrm{res}}$ to the systemic velocity, i.e. by shifting them by $v_{\mathrm{surf}}$, summing:
\begin{equation}
\mathrm{CCF}_{\mathrm{local}}(v) = \sum\limits_{t \in \mathrm{in}}\mathrm{CCF}_{\mathrm{res},-v_{\mathrm{surf}}(t)}(v,t)\ ,
\end{equation}
and then normalising the continuum to unity. We used only the $\mathrm{CCF}_{\mathrm{res}}$ falling between the second and third contact (i.e., fully in transit). We were then able to analyze the profile shape for the local master CCF, including its contrast and FWHM (see Fig.~\ref{ccf_master}). We measure a $\mathrm{FWHM}_{\mathrm{local}}=6.69\pm0.19$ \kms\ and a contrast $C_{\mathrm{local}}=0.505\pm0.012$. In comparison, for the disk-integrated master CCF we measure a $\mathrm{FWHM}=6.747\pm0.008$ \kms\ and a contrast $C=0.503\pm0.001$. Applying the $v \sin i_{*}$ calibration law of \citet{Melo2001} to our two masters, we measure a nonsignificant $v \sin i_{*}=0.7\pm1.2$~\kms. We are thus limited by the $S/N$ ratio and cannot conclude whether the $\widetilde{\mathrm{CCF}}_{\mathrm{local}}$ is intrinsically different from the $\widetilde{\mathrm{CCF}}_\mathrm{out}$. In summary, this analysis does not bring us more information than the analysis of the local stellar surface velocities, but does confirm that the $v \sin i_{*}$ of WASP-49 may remain undetectable.

\subsection{Implications for transmission spectroscopy}\label{subSec_CLB_trans}
\begin{figure}[tbp]
\centering
\includegraphics[width=0.47\textwidth]{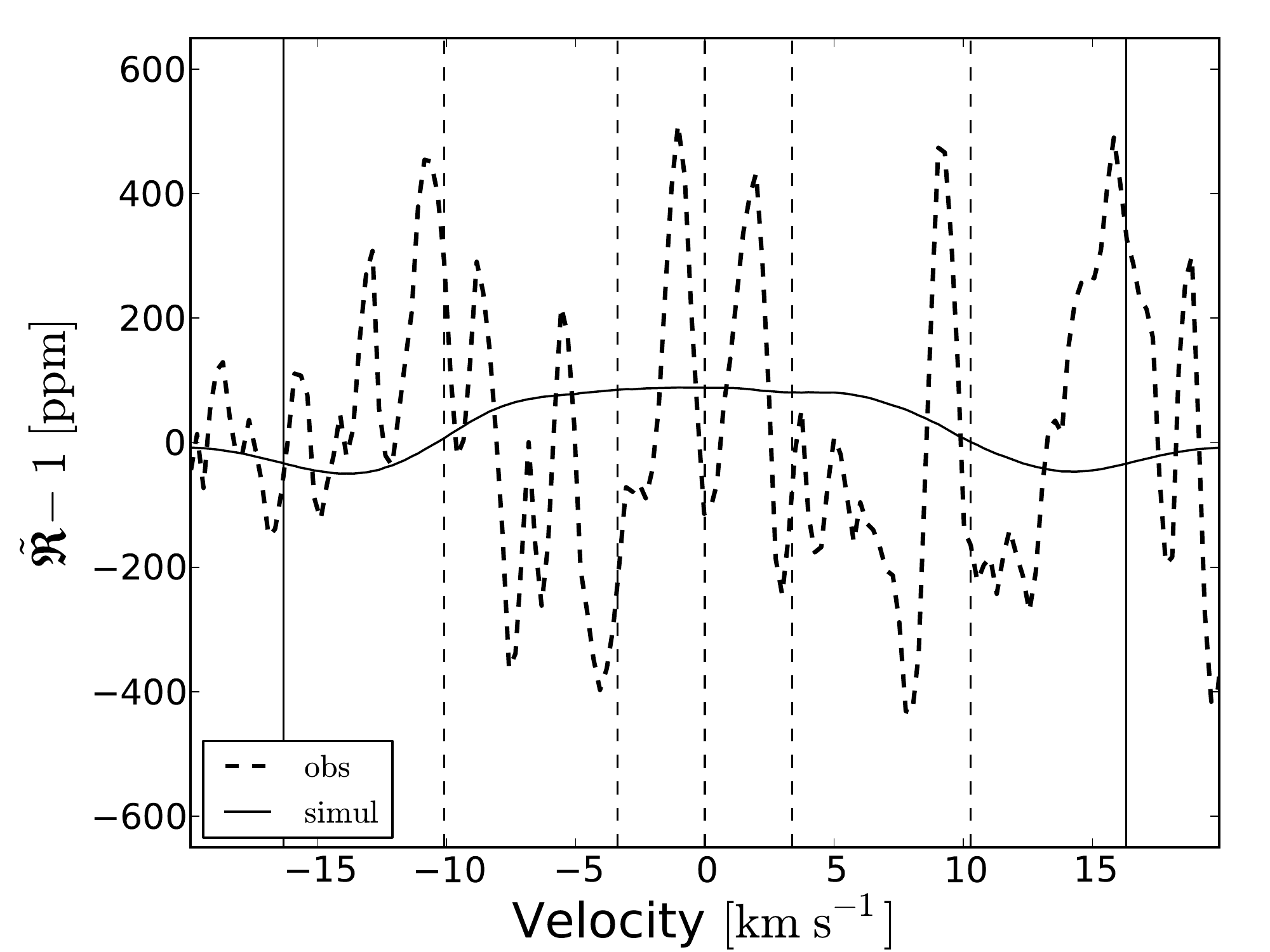}
\caption{Transmission CCF for WASP-49b. This shows the spurious residuals that the Rossiter-McLaughlin effect could imprint on the transmission spectrum at the location of a typical stellar absorption line. The two transmission CCFs shown are computed from the observed data (dashed line) and simulated using the master CCFs (full line). The vertical dashed lines show the limits of the normalization regions (outer two lines at $v\simeq\pm10.1$~\kms), FWHM of the $\widetilde{\mathrm{CCF}}_{\mathrm{out}}$ (middle two lines at $v=\pm3.37$~\kms), $\mathrm{CCF}$ centroid (central line at $v=0$~\kms). The systemic velocity was subtracted. The two outer continuous vertical lines represent the maximum planetary radial velocity during the transit ($v\simeq\mp16.3$~\kms at the first and fourth contact).}
\label{TS_CCF}
\end{figure}

Stellar rotation, through the Rossiter-McLaughlin effect, is known to affect the transmission spectrum \citep{Louden2015,Brogi2016}. We can now examine the shape, position, and amplitude of this effect in our data. In particular, we can compute a ``transmission CCF'' by considering the CCFs as proxies for typical stellar lines (see Appendix~\ref{app:trans_ccf}). The transmission CCF then represents the typical residuals imprinted at the location of each stellar line by the Rossiter-McLaughlin effect. The transmission CCF for WASP-49 is shown in Fig~\ref{TS_CCF}. It exhibits a standard deviation of $\sim200$ ppm and no particular pattern. We also simulated the transmission CCF by assuming the maximum $v\,sin(i_{*})$ allowed by our MCMC analysis ($v \sin i_{*} = 0.46$~\kms) and an aligned spin-orbit configuration. In the case of WASP-49, owing to the very low $v \sin i_{*}$ and the similarity between the local and global master CCFs, we obtain from our simulation that the Rossiter-McLaughlin effect cannot imprint spurious signals larger than $\sim100$ ppm (at 3-$\sigma$). This is below the noise level we have in our transmission spectrum and is thus negligible for the purpose of this paper.

One may ask if the previous discussion concerning CCFs and thus typical stellar lines (such as iron lines) is also applicable to substantially different line shapes such as the \ion{Na}{i} D lines. Generally speaking, the amplitude of Rossiter-McLaughlin residuals in the transmission spectrum scales with the slope of the spectral features, i.e., the narrower and deeper a line, the larger the RM residuals. In this context, we note that the Gaussian line cores of the \ion{Na}{i} D doublet exhibit spectral slopes that are no steeper than the typical iron lines used to build the CCFs. We therefore expect RM residuals in the sodium lines to be no larger than the transmission CCF.

Finally, it is possible that intrinsic center-to-limb variations in the \ion{Na}{i} D doublet line shapes could also induce spurious residuals in the transmission spectrum, as noted by \citet{Czesla2015} (see also \citet{Khalafinejad2016}). However, \citet{Czesla2015} show that center-to-limb variations in strong stellar lines, such as the \ion{Na}{i} D lines, are concentrated in the line wings and may not affect a planetary atmosphere signature lying mostly within the Gaussian line cores. For example, \citet{Khalafinejad2016} used a center-to-limb variation component to model their transmission data, but did not significantly detect it into the line cores. Furthermore, the center-to-limb variations observed by \citet{Czesla2015} and \citet{Khalafinejad2016} within the sodium lines reveal a limb brightening that may cause a decrease of the planetary atmospheric absorption. They also showed that this effect is less important for the earlier spectral types (G stars), which is the case for WASP-49. Given this aspect and the fact that the Gaussian part of the sodium lines is less affected by center-to-limb variations, we argue that this effect is likely negligible in the case of WASP-49. We further strengthen this conclusion below by studying a "control" star of similar spectral type (see Sect.~\ref{sec:alternatives}).

\section{Transmission spectroscopy of WASP-49b}\label{Sec_TransSpec}

\subsection{Extraction of the transit spectrum}\label{Sec_spec_method}
\begin{figure*}[t!]
\centering
\includegraphics[width=0.97\textwidth]{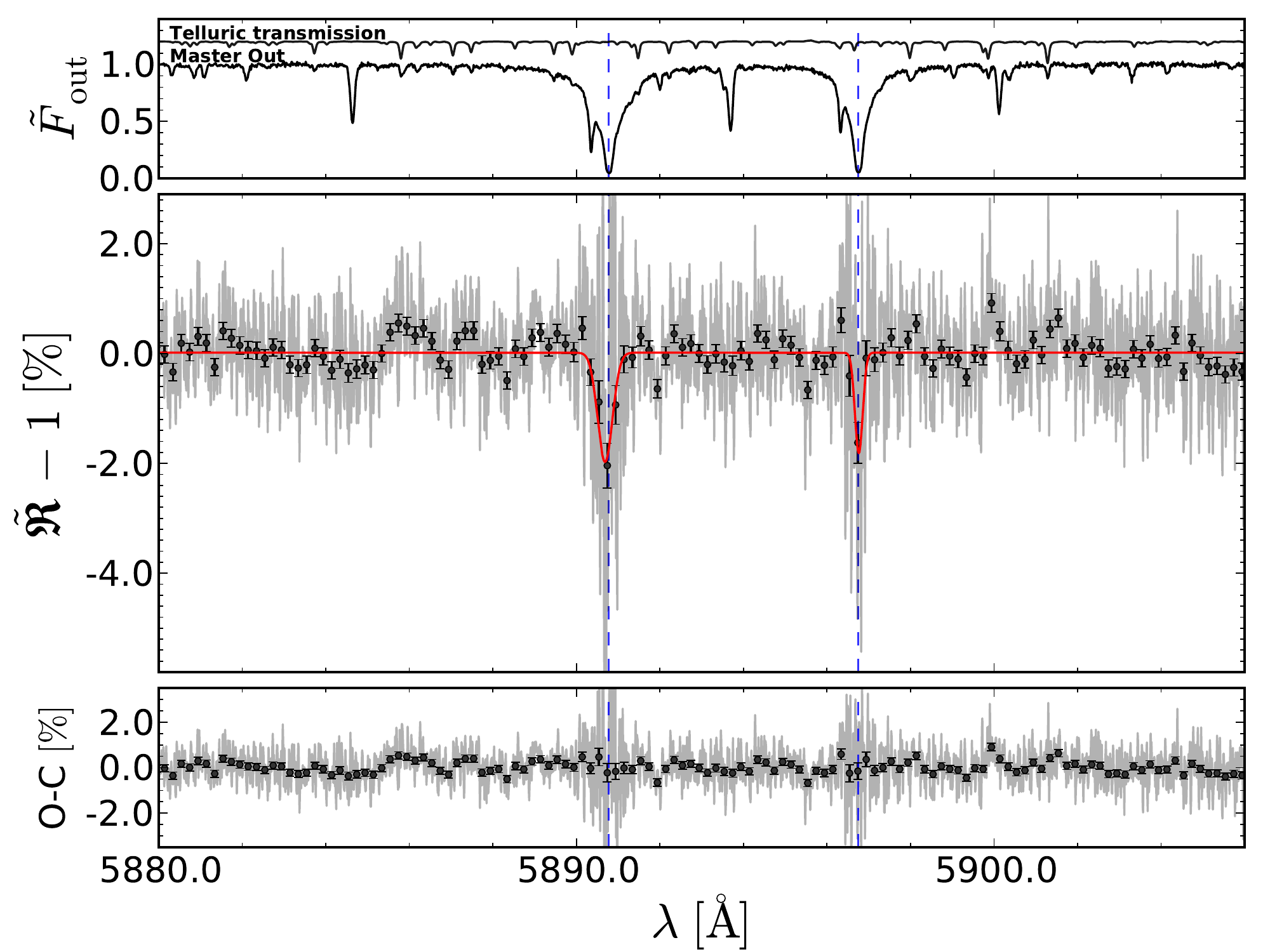}
\caption{Transmission spectrum of WASP-49b in the region of the sodium \ion{Na}{i} D doublet as observed with HARPS. Upper panel: Master-out spectrum and a reference telluric spectrum, for the first night, in the solar system barycentric reference frame. Middle panel: Transmission spectrum of the exoplanet atmosphere (light gray), also binned by 20$\times$ in black circles. We show a Gaussian fit to each sodium line (red) with contrast of 1.99$\pm$0.49\% and 1.83$\pm$0.65\%, and FWHM of 0.42$\pm$0.1 \AA\ and 0.22$\pm$0.08 \AA\ for the D$_2$ and D$_1$ line, respectively. The rest-frame transition wavelengths are shown with the blue dashed line. Lower panel: Residuals to the Gaussian fit.}
\label{fig:TransitSpectrum}
\end{figure*}

\begin{figure*}[htbp]
\centering
\includegraphics[width=0.47\textwidth]{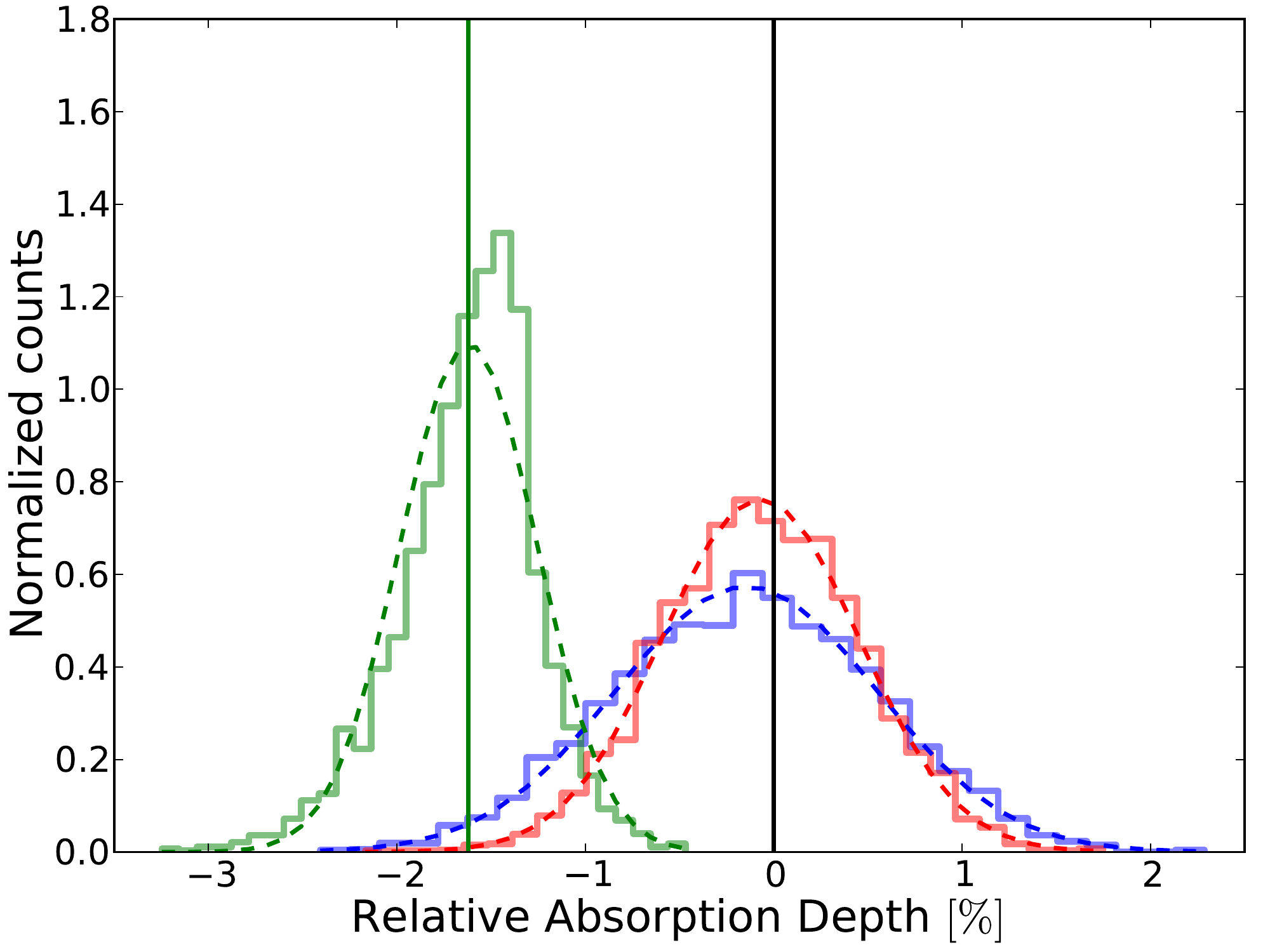}
\includegraphics[width=0.47\textwidth]{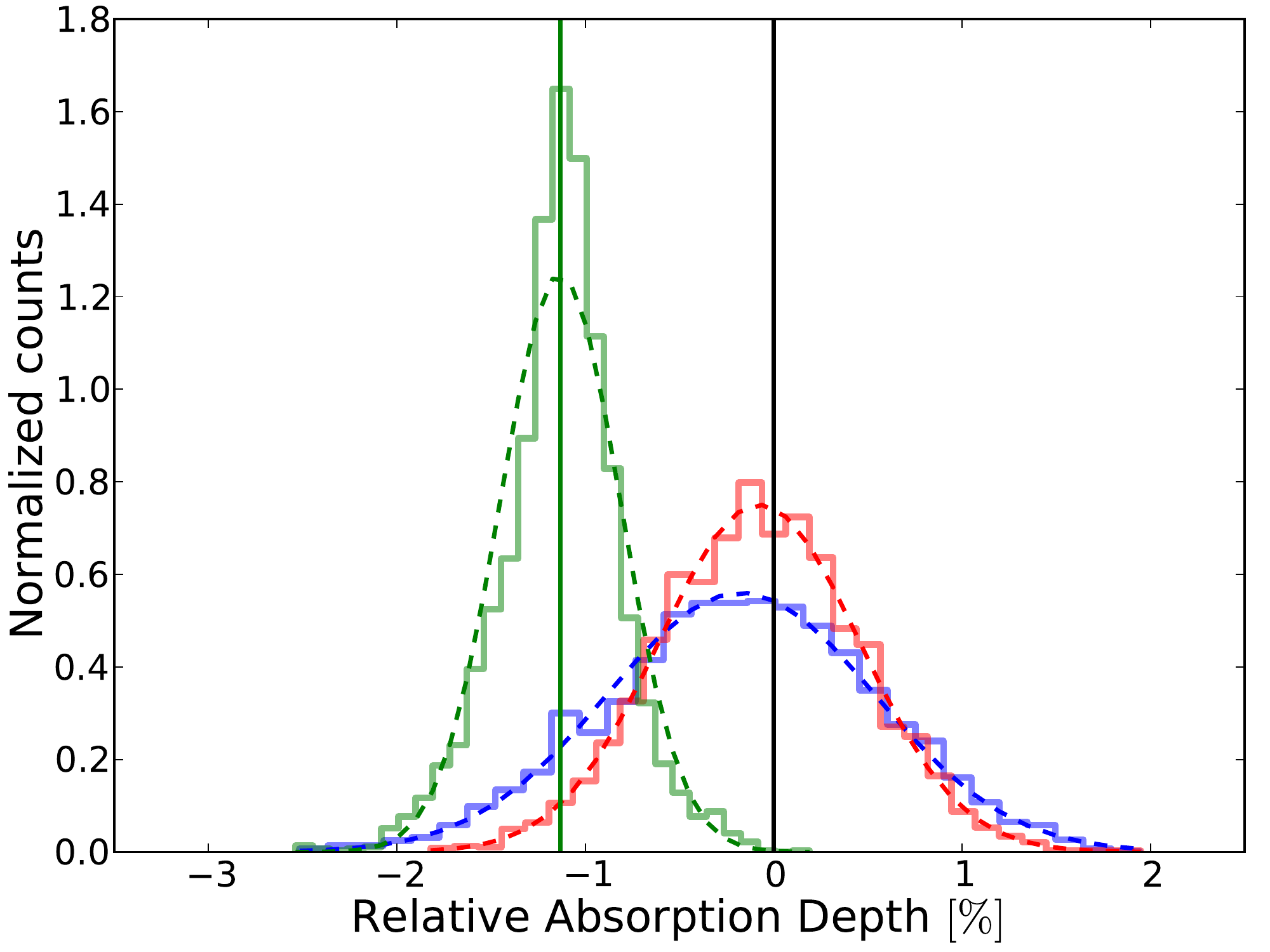}
\caption{Distribution obtained with the empirical Monte Carlo analysis for the \ion{Na}{i} D$_2$ line (left) and D$_1$ line (right). In-in, out-out, and in-out distribution (continuous lines) are shown in blue, red, and green with their respective Gaussian fit (dotted lines). The green vertical lines are the averages values of the in-out distribution.}
\label{EMC}
\end{figure*}

Ground-based transit spectroscopy is a double-differential technique. As for space-borne measurements, a temporal differentiation is needed: stellar spectra recorded during transit $f(\lambda,t_\mathrm{in})$, which contain the transmission signature of the planetary atmosphere, are compared to the stellar spectra obtained out of transit $f(\lambda,t_\mathrm{out})$. 
These spectra are first corrected for telluric signatures as described in Sect.~\ref{subSec_tellcorr}. They are then stacked in and out of transit to obtain the reference master-in and master-out spectra, $F_\mathrm{in}(\lambda)$ and $F_\mathrm{out}(\lambda)$, respectively. These master spectra are classically divided to produce the transit spectrum, or spectral ratio, $\mathfrak{R} = F_\mathrm{in}/F_\mathrm{out}$ \citep{Brown2001}.

Meanwhile, our ground-based observations first require a spectral differentiation, before the temporal differentiation can be performed. This is because of the time-correlated noises introduced by variations in the Earth atmosphere. This normalization is carried out by dividing the stellar spectra  by their respective values in a reference wavelength band $\langle\lambda_\mathrm{ref}\rangle$, chosen in the continuum, far away from the \ion{Na}{i} line cores. The normalized in and out spectra are
\begin{eqnarray}
\tilde{f}(\lambda, t_\mathrm{in}) & = & \frac{f(\lambda,t_\mathrm{in})}{f(\langle\lambda_\mathrm{ref}\rangle,t_\mathrm{in})}, \\
\tilde{f}(\lambda, t_\mathrm{out}) & = & \frac{f(\lambda,t_\mathrm{out})}{f(\langle\lambda_\mathrm{ref}\rangle,t_\mathrm{out})}.
\end{eqnarray}
The master spectra are similarly self-normalized, yielding $\tilde{F}_\mathrm{in}(\lambda)$ and $\tilde{F}_\mathrm{out}(\lambda)$. This spectral normalization preserves the signatures of the planetary atmosphere appearing during the transit; however, this is achieved at the cost of correcting out for the ``white light'' transit signal. Thus, these observations cannot provide any constraints on the planet broadband radius. They can only reveal radius variations from differential absorption signatures in the planet atmosphere.

Once the spectra are corrected for tellurics and normalized, we apply the temporal differentiation between the spectra taken in- and out-of-transit to yield the normalized transit spectrum, $\tilde{\mathfrak{R}}$. We emphasize, however, that $\tilde{\mathfrak{R}} \neq \tilde{F}_\mathrm{in}/\tilde{F}_\mathrm{out}$ because of the planet motion during the transit, a spectral ratio $\tilde{f}/\tilde{F}_\mathrm{out}$ has to be calculated at each time step in transit and shifted in wavelength to the planet rest frame. Doing otherwise would cause the dilution of the planet atmospheric signal. Therefore, following \citet{Wyttenbach2015}, we extract the transit spectrum by coadding the shifted spectral ratios obtained at each time step during the transit, 
\begin{equation} \label{eq:TransitSpectrum}
\tilde{\mathfrak{R}}(\lambda) = \sum\limits_{t \in \mathrm{in}} \left. \frac{\tilde{f}(\lambda,t)}{\tilde{F}_\mathrm{out}(\lambda)}\right|_{p}.
\end{equation}
We verified that the wavelength shift does not affect the shape of the spectral ratios or add extra noise. The transit spectrum of WASP-49b is plotted in Fig.~\ref{fig:TransitSpectrum}.

\subsection{The sodium signature}\label{Sec_SodiumD}

A clear double-peaked signal appears in the transit spectrum of WASP-49b (Fig.~\ref{fig:TransitSpectrum}). The features are centered at wavelengths close to the sodium doublet lines D$_1$ and D$_2$, at $5895.924$~\AA\ and $5889.951$~\AA, respectively. Since WASP-49 has a systemic velocity $\gamma = 41.7$~\kms, these lines can be observed at $5890.771$~\AA\ (D$_2$) and $5896.745$~\AA\ (D$_1$) in the solar system barycentric reference frame. Thanks to the relatively high value of $\gamma$, the stellar sodium lines are well separated from the interstellar medium lines (at $\sim$20~\kms, see Fig.~\ref{fig:TransitSpectrum}) and from the telluric sodium lines (located around the BERV values of $6.6$, $-3.5$, and $-9$~\kms during the different nights of observation).

First, we assess the statistical significance of the signal and ensure it arises from the planetary transit. For this, we use the signal absorption depth $\delta$ as a figure of merit: $\delta$ is calculated as the weighted mean of $\tilde{\mathfrak{R}}$ over a passband $C$, composed of two sub-bands centered on the core of each feature. Our tests have shown that a width of 0.4~\AA\ for each sub-band maximizes the signal-to-noise ratio. The signal calculated over $C$ is compared to the average of two reference bands $B = \left[5874.94;5886.94\right]$~\AA\ and $R = \left[5898.94;5910.94\right]$~\AA, taken in the continuum, on the blue and red sides of $C$, respectively. We checked that the choice of the reference bands has no impact on the result. The absorption depth is expressed as
\begin{equation}
\delta = \frac{\sum_{C} w_i \tilde{\mathfrak{R}}(\lambda_i)}{\sum_{C} w_i} - \frac{1}{2} \left(\frac{\sum_{B} w_i \tilde{\mathfrak{R}}(\lambda_i)}{\sum_{B} w_i} + \frac{\sum_{R} w_i \tilde{\mathfrak{R}}(\lambda_i)}{\sum_{R} w_i}\right),
\end{equation}
where the weights are the inverse of the squared uncertainties on $\tilde{\mathfrak{R}}$, $w_i = 1/\sigma_i^2$. The uncertainties are assumed to arise from photon and readout noise and are propagated from the individual spectra.

Fixing the $C$, $B$, and $R$ bands, we calculate the distribution of $\delta$ values obtained through an Empirical Monte-Carlo (EMC) simulation \citep{Redfield2008,Astudillo-Defru2013,Wyttenbach2015}. This is achieved by randomizing the time tags of the stellar spectra $f$ in transit and by selecting a subsample of spectra in transit. The latter is compared to the $\tilde{F}_\mathrm{out}$ through Eq.~(\ref{eq:TransitSpectrum}). We obtain a set of transit spectra out of which an ``in-out'' distribution of $\delta$ values is obtained. As in \citet{Wyttenbach2015}, we also use two control distributions, ``out-out'' and ``in-in'', constructed by randomly selecting and comparing spectra $f$ only in the out-of-transit sample or only in the in-transit sample, respectively. These control distributions should bear no transit signal and we verify that they are properly centered on 0 for both sodium lines in Fig.~\ref{EMC}. We find that the ``in-out'' distribution of absorption depths is significantly different from the control distributions, for both lines (together or individually) and during each observation night. The results are summarized in Table~\ref{table3}. Coadding the three observation nights, the extra absorption signal in the \ion{Na}{i} D$_1$ and D$_2$ lines are measured at $1.11\pm0.25\%$ (4.4$\sigma$) and $1.44\pm0.28\%$ (5.2$\sigma$), respectively. Considering both lines together, we detect sodium in the transmission spectrum of WASP-49b with an absorption depth of  $1.26\pm0.19\%$ (6.8$\sigma$). The EMC also reveals a significant absorption of $1.38\pm0.31$ (4.5$\sigma$). The uncertainty on the latter value captures systematic noise in addition to the photon noise assumed to be the sole contributor to the uncertainty on the earlier value. This shows that the global error budget is 1.5$\times$ the photon noise. In the following, we use the more conservative value from the EMC analysis.

\subsection{Possible origins of the sodium signature}
\label{sec:alternatives}

The signature of neutral sodium has been expected \citep{Seager2000} and detected multiple times \citep{Charbonneau2002,Wyttenbach2015,Sing2016} in the atmospheres of hot gas giants. Yet, the strength of the signature we observe on WASP-49b is outstanding in several respects, as discussed below in Sect.~\ref{sec:atmosphere}. Therefore, before studying implications for the exoplanet atmosphere, we examine possible alternative origins for this signature.

Several authors have closely examined the methodology and results of \citet{Wyttenbach2015}, who have used similar observations, reduction, and analysis techniques as presented here, although applied to the much brighter (and more active) HD~189733b. \citet{Louden2015} and \citet{Brogi2016} drew attention to the treatment of the Rossiter-McLaughlin effect, which could affect the shape of spectral features in the transit spectrum if it is not properly accounted for. The present analysis makes use of the new approach developed by \citet{Cegla2016b} to analyze the Rossiter-McLaughlin effect. This is discussed at length in Sect.~\ref{Sec_CLB}. We demonstrate in Sect.~\ref{Sec_CLB} that WASP-49 is a very slow rotator (or is seen pole on), which largely tames the impact of the Rossiter-McLaughlin effect on our data.

Furthermore, \citet{Barnes2016} cautioned that stellar activity could create spurious signatures in lines originating mainly from the stellar chromosphere, in particular, in the \ion{Ca}{ii} H and K and H$\alpha$ lines. According to these authors, the \ion{Na}{i} doublet lines whose narrow cores are formed close to the chromosphere could also be sensitive to such effects, although to a lesser extent than the lines quoted above. The star WASP-49, however, is much less active than HD~189733. In fact, WASP-49 has an extremely low $\log R'_{HK}$ of $-5.17$ compared to HD~189733 \citep[$\log R'_{HK} \simeq -4.5$;][]{Cegla2016b}.

\begin{figure*}[htbp]
\centering
\includegraphics[width=0.33\textwidth]{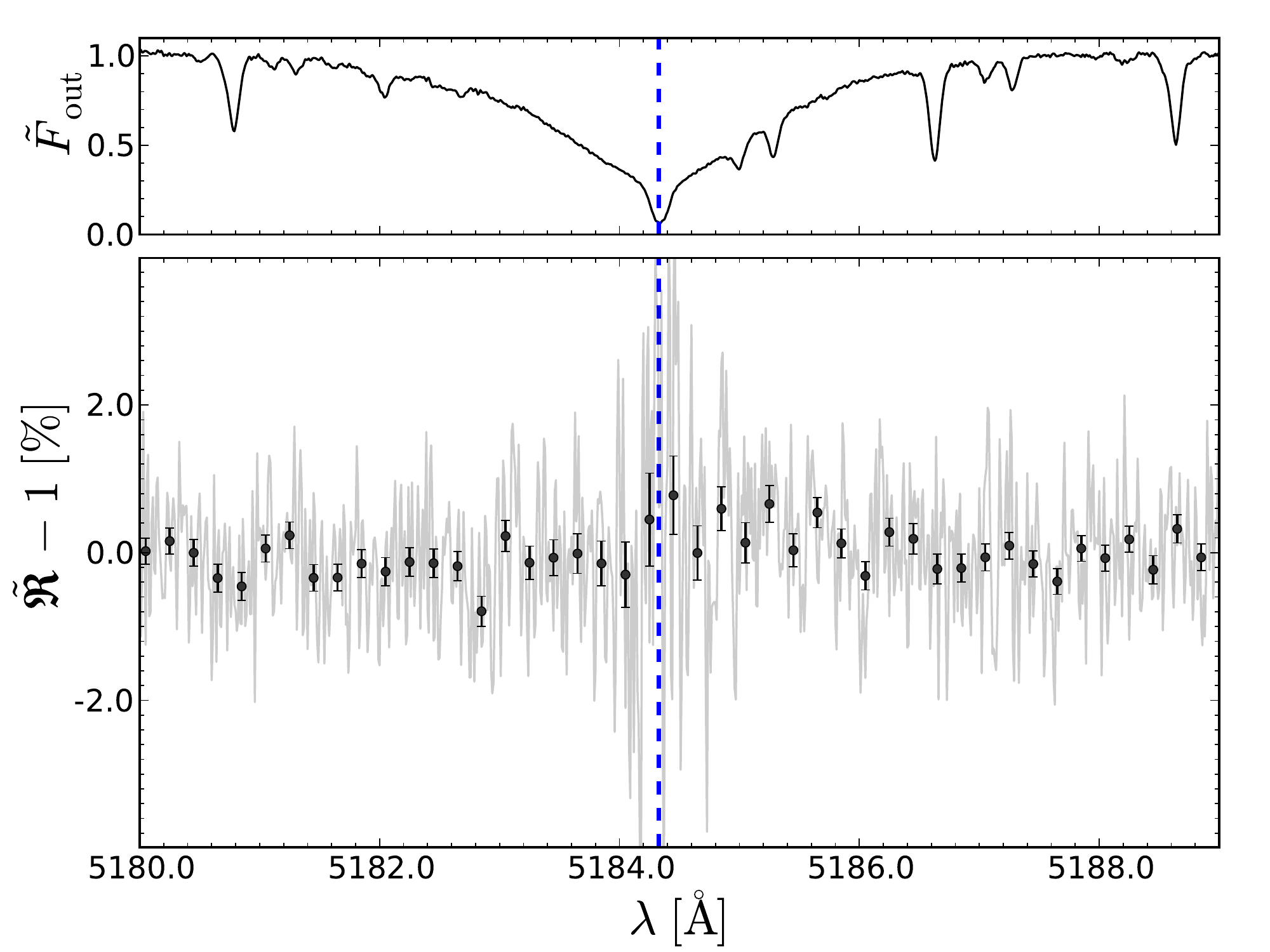}
\includegraphics[width=0.33\textwidth]{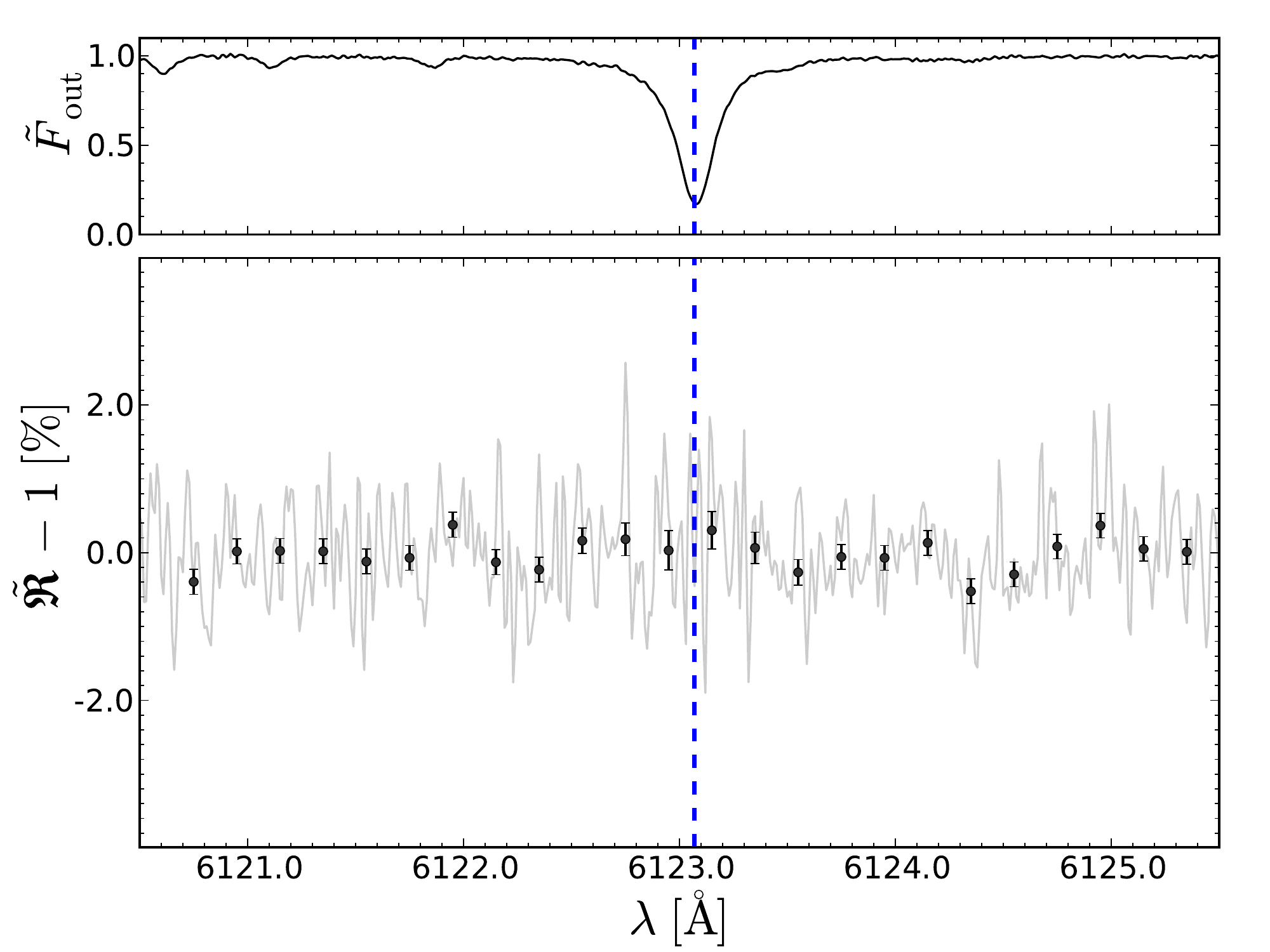}
\includegraphics[width=0.33\textwidth]{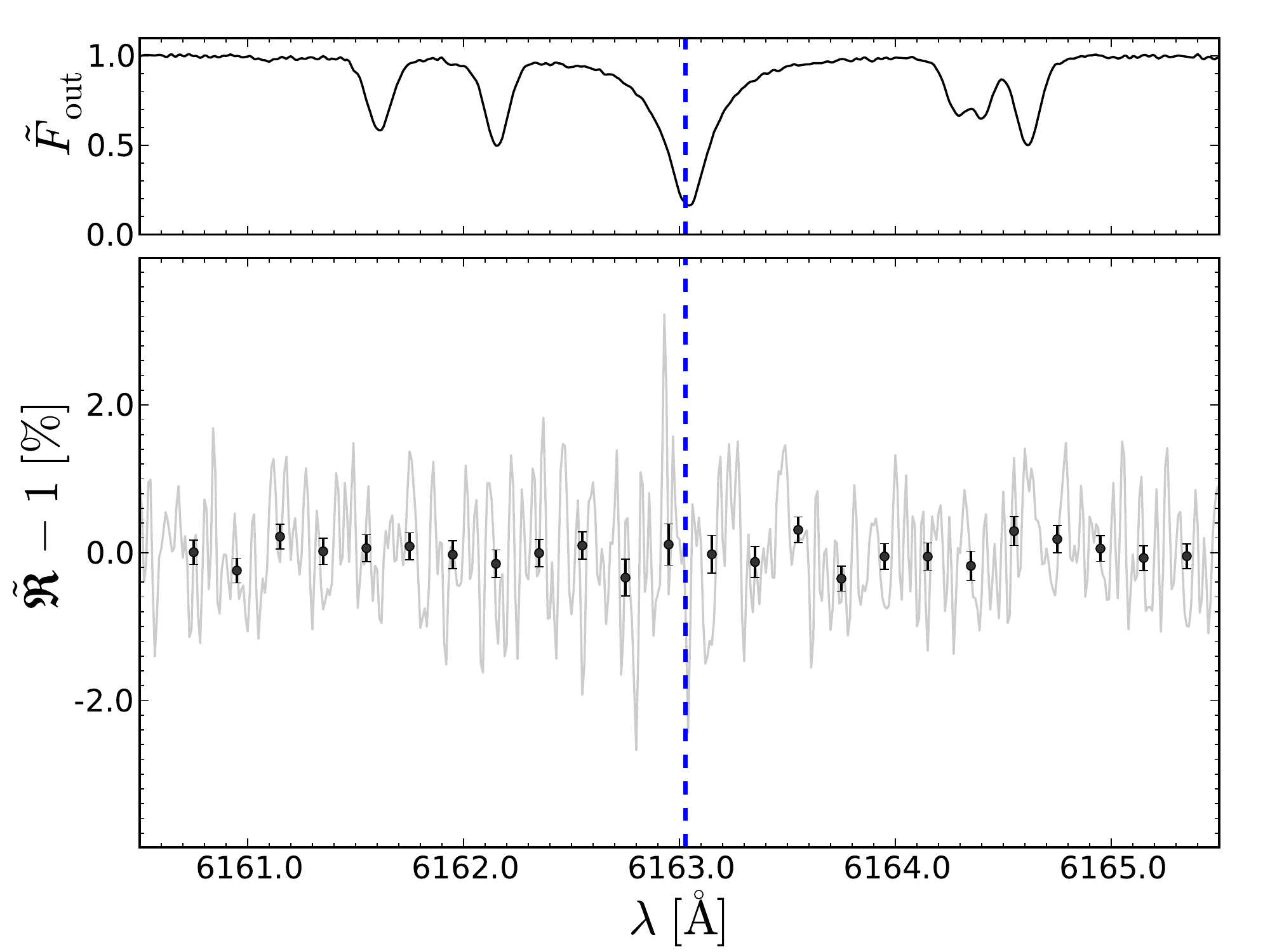}
\caption{Transmission spectrum around the \ion{Mg}{i} $\lambda$5183.604 (Fraunhofer b) and \ion{Ca}{i} at 6122.217~\AA\ and 6162.173~\AA\ control lines. No spectral features are detected.}
\label{TS_control}
\end{figure*}

As control experiments, we derived the transmission spectrum of WASP-49b around other strong stellar lines in the HARPS spectra, namely \ion{Mg}{i} at 5183.604~\AA\ (the Fraunhofer b line) and \ion{Ca}{i} at 6122.217~\AA\ and 6162.173~\AA. The resulting transit spectra are shown in Fig.~\ref{TS_control}. The EMC simulations allow us to measure absorption depth in a 0.4$\AA$ passband, for the three lines, of $0.26\pm0.53\%$, $0.28\pm0.24\%$ and $0.0\pm0.28\%$, respectively. No signatures are detected, in striking contrast with the sodium lines, showing that the sodium excess absorption signal does not result from a systematic effect in our analysis procedure, nor from a chromospheric activity contribution.

\begin{figure}[t!]
\centering
\includegraphics[width=0.47\textwidth]{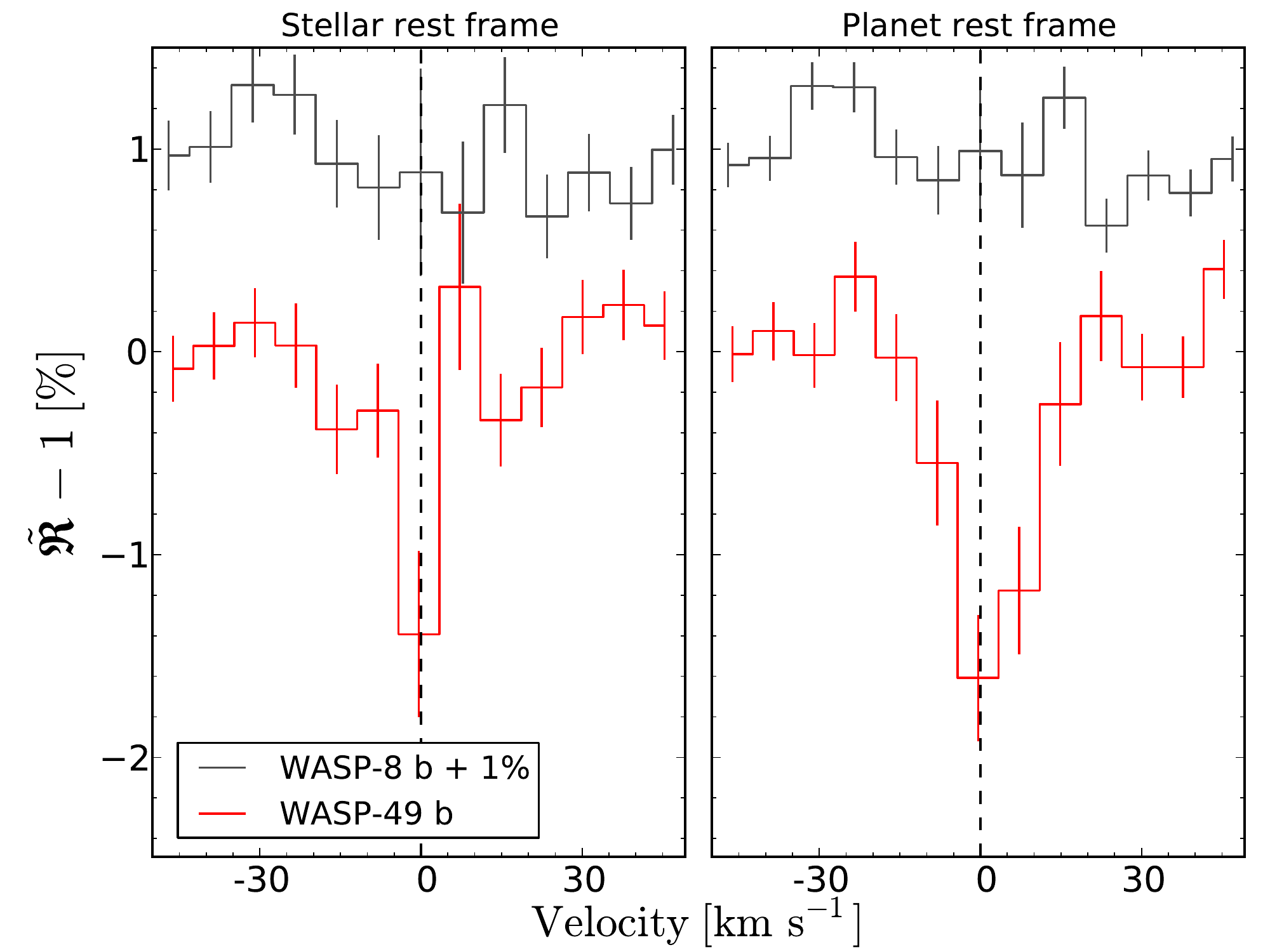}
\caption{Transmission spectrum of the coadded sodium D$_1$ + D$_2$ shown in velocity space for WASP-8b (acting as a control star, in black) and WASP-49b (in red), both binned by $15\times$. A sodium absorption is present in both rest frames for WASP-49b, but slightly more significant in the planet rest frame, as expected for a planetary signature. Both stars are classified G6V and share common properties, except that WASP-8b is much more active, which could possibly lead to spurious sodium signatures of stellar origin. However, no sodium signal can be seen in WASP-8b in spite of its high activity level. Thus it is not likely that the sodium signal of WASP-49 has a stellar activity origin.}
\label{fig:TransitSpectrumW8}
\end{figure}

\citet{Barnes2016} claims that the effect of stellar activity can be seen in the transmission spectrum computed in the stellar rest frame, i.e., by not correcting the planet velocity in Equation~\ref{eq:TransitSpectrum}. Despite the low activity index of WASP-49, we verified this affirmation by comparing the significance of the sodium detection in both the stellar and planet rest frame (see Table~\ref{tab:GaussianFit} and Fig.~\ref{fig:TransitSpectrumW8}). We note a global consistency of all derived line parameters between the \ion{Na}{i} D$_1$ and D$_2$ lines, on the one hand, and the stellar and planetary frames on the other hand. There is a mild tension with the D$_1$ line in the stellar rest frame, which appears deeper and narrower than the other features, however, it has an integrated absorption depth (on a $0.4~\AA$ passband) that is not significantly detected. When looking at the coadded sodium lines in the two different rest frames, the absorption depths $\delta_{\rm EMC}$ (computed on a $2\times0.4~\AA$ passband) appear stronger in the planet rest frame, which supports a planetary origin.

\begin{table}[t!]
\caption{Summary of the measured relative absorption depth $\delta$ in [$\%$] of the transmission spectrum of WASP-49b. The integration passband is $\Delta\lambda=0.4~\AA$ for each \ion{Na}{i} line ($\sim25$ pixels). The ``photon'' labels show the values measured on our nominal spectra together with the propagated internal uncertainties. The ``EMC'' labels represent the values and errors given by the EMC distributions.}
\begin{center}
\begin{tabular}{lccc}
\hline
\rule[0mm]{0mm}{5mm}\ion{Na}{i} line(s) & D$_2$&D$_1$&D doublet\\
\hline
Night 1 (Photon) &2.99$\pm$0.47&1.55$\pm$0.43&2.21$\pm$0.32\rule[0mm]{0mm}{3mm}\\
Night 1 (EMC) &2.84$\pm$0.73&1.00$\pm$0.62&1.76.$\pm$0.47\\
Night 2 (Photon) &1.22$\pm$0.59&1.93$\pm$0.55&1.60$\pm$0.40\\
Night 2 (EMC) &1.16$\pm$0.78&2.28$\pm$0.78&1.72$\pm$0.55\\
Night 3 (Photon) &1.63$\pm$0.38&1.25$\pm$0.35&1.42$\pm$0.26\\
Night 3 (EMC) &2.24$\pm$0.54&1.52$\pm$0.55&1.89$\pm$0.39\\
\hline
All Nights (Photon) &1.44$\pm$0.28&1.11$\pm$0.25&1.26$\pm$0.19\\
All Nights (EMC) &1.62$\pm$0.44&1.13$\pm$0.44&1.38$\pm$0.31\\
\hline
\end{tabular}
\end{center}
\label{table3}
\end{table}

We proceed with an additional control experiment: the analysis of archival HARPS data of the transiting gas giant WASP-8b \citep[ESO program 082.C-0040;][]{Queloz2010,Bourrier2016}. The data set is composed by one night of observation, comprising 75 spectra, out of which 47 are in transit. The data are of good quality (S/N range from 31 to 56) and the transit is well covered, allowing us to compute a good quality transmission spectrum for WASP-8b. WASP-8b is $\sim$5.5$\times$ more massive than WASP-49b and has a $\sim$10$\times$ smaller atmospheric scale height; therefore no atmospheric signatures are expected to be detectable. WASP-49 and WASP-8 are both G6V stars, making the WASP-8 spectrum (stellar lines and CCF) a close analog to WASP-49. From the available spectra, we calculate a value of $\log R'_{HK} = -4.62\pm0.01$ for WASP-8, making it much more chromospherically active than WASP-49. If the observed signal on WASP-49 was arising from stellar activity, a similar analysis on WASP-8 should arguably yield a similar, if not enhanced, signature. However, the resulting WASP-8b transit spectrum around the region of \ion{Na}{i} (shown in Fig.~\ref{fig:TransitSpectrumW8}) is flat. No sodium absorption signal is measured in any of the \ion{Na}{i} D lines\footnote{The impact of the RM effect due to the WASP-8b misaligned spin-orbit on the transmission spectrum ($\sim0.1\%$) is negligible in comparison to the data noise in the transmission spectrum ($\sim0.4\%$).}, setting an upper limit on a total sodium absorption signal in WASP-8b of $<0.54\%$ (3$\sigma$; EMC results in a $2\times0.4$~\AA\ passband). This experiment on a control system further strengthens the planetary nature of the signal detected during the transit of WASP-49b.

\begin{table}[t!]
\caption{Summary of the Gaussian fit parameters and absorption depths for WASP-49b in the planetary and stellar rest frames}
\begin{center}
\begin{tabular}{lccc}
\hline
\rule[0mm]{0mm}{5mm}\ion{Na}{i} line(s) & D$_2$&D$_1$&D$_1$ + D$_2$\\
\hline
\leavevmode\\
\multicolumn{2}{l}{Planet frame:}\leavevmode\\
Contrast [$\%$] &1.99$\pm$0.49&1.83$\pm$0.65&1.87$\pm$0.41\rule[0mm]{0mm}{3mm}\\
FWHM [\kms]&21.3$\pm$4.9&11.1$\pm$4.3&15.0$\pm$3.6\\
$\mu$ [\kms] &-4.3$\pm$4.8&0.8$\pm$4.2&-1.7$\pm$1.6\\
$\delta_{\rm EMC}$ [$\%$] &1.62$\pm$0.44&1.13$\pm$0.44&1.38$\pm$0.31\\
(S/N)$_\delta$ &3.7&2.6&4.5\\
\leavevmode\\
\multicolumn{2}{l}{Stellar frame:}\leavevmode\\
Contrast [$\%$] &2.69$\pm$0.80&8.33$\pm$1.92&3.85$\pm$0.85\rule[0mm]{0mm}{3mm}\\
FWHM [\kms] & 16.1$\pm$3.5&4.8$\pm$1.2&6.9$\pm$1.6\\
$\mu$ [\kms] &-0.8$\pm$1.4&0.1$\pm$0.4&0.5$\pm$0.4\\
$\delta_{\rm EMC}$ [$\%$] &1.38$\pm$0.38&0.10$\pm$0.36&0.71$\pm$0.26\\
(S/N)$_\delta$ &3.6&0.3&2.7\\
\hline
\end{tabular}
\end{center}
\label{tab:GaussianFit}
\end{table}

\subsection{Neutral sodium in WASP-49b thermosphere}\label{sec:atmosphere}

As in the case of HD~189733b \citep{Wyttenbach2015}, we interpret the observed sodium signature in WASP-49b as absorption at the atmospheric planetary limb. We first characterize the spectroscopic features by fitting a Gaussian to each sodium line (see Fig.~\ref{fig:TransitSpectrum}). The Gaussian parameters (contrast, FWHM, and mean) are provided in Table~\ref{tab:GaussianFit} for both lines individually (D$_2$ and D$_1$) and coadded ($\rm D_1+D_2$). Both lines have a similar contrast ($\lesssim2\sigma$), as expected in the case of a transit spectrum.\footnote{In the case of an isothermal atmosphere, the contrast difference between both lines is of order $H \ln 2$, which is a too small value to be detected with these observations.} The contrast of the coadded sodium doublet is $1.79\pm0.36\%$, which is 3.4$\times$ larger than for HD~189733b \citep{Wyttenbach2015}, while {the expected transmission signal} ratio of the two planets is only of $\sim$2.3. Assuming the planetary radius of \citet{Lendl2016}, the absorption is $\sim80\times$ the absorption of one atmospheric scale height. This implies that the sodium absorption signature must arise from a very high altitude in the planetary atmosphere, at $\sim$1.5 planet radius. This corresponds to an equivalent optically thick layer of $\sim$45,000~km. The FWHM of the coadded sodium line is $14.8\pm3.3$~\kms. The line profile is thus resolved by HARPS (with a 2.7~\kms\ resolution element) by a factor of $\sim$5.5. In contrast with HD~189733b \citep{Wyttenbach2015,Louden2015}, we do not detect a significant (i.e., $>3\sigma$) velocity shift of the line centers. 

When spectral lines are resolved, it is in principle possible to adjust different parts of the lines, which originate from different parts of the atmosphere, with isothermal models, and reconstruct from this the temperature profile of the atmosphere, as in HD~189733b \citep{Wyttenbach2015}. However, the signal-to-noise ratio of WASP-49b transit spectrum is lower than for HD~189733b (owing to the 3.6~mag difference between both stars), making this method difficult to apply.

\begin{figure*}[t!]
\centering
\includegraphics[width=0.97\textwidth]{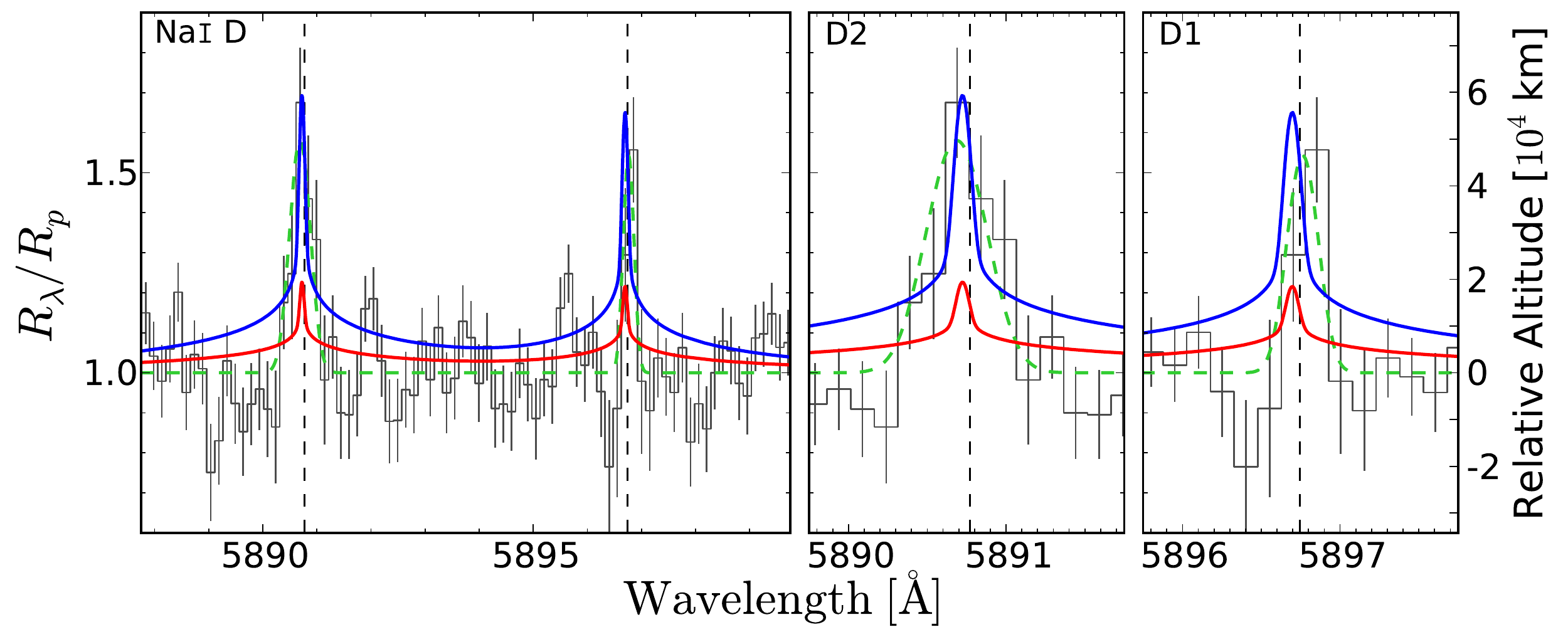}
\caption{Fit of $\eta$ models to the transmission spectrum of WASP-49b. The spectra are binned by $15\times$. The vertical scale are relative radii and altitudes assuming a white light radius of $1.198\,R_{\mathrm{Jup}}$ (Table~\ref{tab:w49}). One $\eta$ model, set at the equilibrium temperature ($\sim1,400$~K, in red), is adjusted to the continuum. Another model is adjusted to the line cores and is shown in blue. The latter model temperature is found to be $2,950^{+400}_{-500}$~K. The Gaussian fits are also shown in green for comparison.}
\label{fig:TS_fit}
\end{figure*}

Meanwhile, the data clearly show that high temperatures, far in excess of the equilibrium temperature ($\sim$1,400~K), are required to explain the large contrast measured in the sodium lines. We generated a grid of forward models spanning a temperature range of $1,000$--$4,000$~K with the $\eta$ code \citep{Ehrenreich2006}, and compared them with the data. Each model is consistent with photometry in the wideband filter used by \citet{Lendl2016}. We computed the $\chi^2$ and BIC associated with each model by restricting the wavelength range to two 0.3~$\mathrm{\AA}$-wide bands (each band being equivalent to one fitted Gaussian FWHM) centered in the D$_1$ and D$_2$ line cores. The models with the lowest BIC value (within $\rm \Delta BIC = 5$) best reproduce the data and span the temperature range $2,450$--$3,350$~K, with the best-fit model ($T\sim2,950$~K) shown in blue in Fig.~\ref{fig:TS_fit}. The equilibrium temperature model ($T=1,400$~K, red curve in Fig.~\ref{fig:TS_fit}) evidently fails to reproduce the line contrast ($\rm \Delta BIC\simeq25$).

With a zero-albedo equilibrium temperature of $1,400$~K, WASP-49b is expected to be cloud free \citep{Heng2016}. The pressure scale height, assuming that the temperature is the equilibrium temperature, is about $H = 730$~km. Following the procedure outlined in \citet{Heng2016}, we estimate the cloudiness index associated with the Gaussian fit on the D$_1$ and D$_2$ lines to be $C= 0.22 \pm 0.07$ and $C=0.14 \pm 0.03$, respectively. Obtaining $C<1$ is somewhat odd because the cloudiness index has a theoretical lower limit of unity. A perfectly cloud-free atmosphere has $C=1$ and cloudy atmospheres have $C>1$. Having $C<1$ is probably an artifact of an assumption built into the construction of the cloudiness index, which is that the atmosphere is isothermal. Notwithstanding, our estimates of C indicate that WASP-49b has a cloud free atmosphere at the altitudes probed by the sodium lines. The number densities of sodium probed by the D$_1$ and D$_2$ line centers are $n_0 \sim 10^2$ cm$^{-3}$.

These observations are thus sensitive to the presence of neutral sodium at altitudes where the atmosphere is substantially hotter than the planet equilibrium temperature. This can be understood if the sodium line cores probe the region where high-energy radiation from the star is absorbed, resulting in upper atmospheric heating \citep{Lammer2003,Lecavelier2004,Koskinen2013a,Koskinen2013b}. Therefore, we are probing the planet thermosphere. Our new detection broadens the general understanding of the thermospheric region and of the transition region where the atmospheric escape, seen in hot Jupiters \citep{VidalMadjar2003,Lecavelier2012} and in a warm Neptune \citep{Ehrenreich2015}, takes place \citep[see also][]{VidalMadjar2013}.

\section{Conclusions}\label{Sec_Conclu}

We have collected and analyzed 126 spectra of WASP-49, covering three transits of its low-density, highly irradiated giant planet WASP-49b. The high spectral resolution and stability of the HARPS spectrograph allowed us to obtain the first resolved transmission spectrum of the planet atmosphere. The spectrum is dominated by the sodium doublet, which produces extra transit absorption signatures of $1.99\pm0.49\%$ and $1.83\pm0.65\%$ in the D$_2$ and D$_1$ lines, respectively. This corresponds to a high-altitude absorption signal caused by neutral sodium atoms. Modeling the absorption signal in the cores of the resolved lines requires elevated temperatures of $2,950^{+400}_{-500}$~K, showing that the signal arises from a hotter upper region of the planet atmosphere, which we identify as the thermosphere.

Additionally, the contrast of the sodium signature in WASP-49b is larger than in HD~189733b, hinting at a possible link with the higher level of stellar irradiation the planet receive ($\sim$2$\times$ the level of HD~189733b). WASP-49b is also 3$\times$ lighter than HD~189733b, and thus much closer to edge of the dearth of highly irradiated, intermediate-mass planets \citep{Mazeh2016}. This desert, even more obvious in the mass-incident flux plane (Fig.~\ref{fig:dearth}), is at least partially attributed to atmospheric escape \citep{Lecavelier2007,Davis2009,Ehrenreich2011}. It is thus tantalizing to consider enhanced thermospheric signatures, such as those detected in WASP-49b, as a proxy of the evaporation status of giant exoplanets. 

In addition, the HARPS spectra show that the host star WASP-49 has a very low chromospheric activity index ($\log R'_{HK} = -5.17$) and its slow apparent rotation (the star might be seen pole on) is unlikely to create spurious transit signatures. It has been suggested that a low value of $\log R'_{HK}$ in a hot gas giant system could be explained by a torus of gas replenished by the planet, which would damp the \ion{Ca}{ii} line core emission \citep{Fossati2013}. However, we lack the elements to constrain this scenario.

New observations at high spectral resolution of a sample of giant exoplanets in different irradiation conditions will substantially increase our understanding of the atmospheric response to a tremendous incoming flux. The survey we have undertaken with HARPS at a 4-meter-class telescope is a pathfinder toward what will be possible to achieve with the upcoming ESPRESSO spectrograph, which will be installed in 2017 at the 8-meter-class Very Large Telescope.

\begin{acknowledgements}
This work has been carried out within the frame of the National Centre for Competence in Research ``PlanetS'' supported by the Swiss National Science Foundation (SNSF). The authors also acknowledge the continuous financial support of the SNSF by the grant numbers 200020\_152721 and 200020\_166227. This publication makes use of The Data $\&$ Analysis Center for Exoplanets (DACE), which is a facility based at the University of Geneva (CH) dedicated to extrasolar planets data visualization, exchange, and analysis. DACE is a platform of the Swiss National Centre of Competence in Research (NCCR) PlanetS, federating the Swiss expertise in Exoplanet research. The DACE platform is available at https://dace.unige.ch. We thank the anonymous referee for the careful reading and pertinent comments. We thank M.~Lendl, D.~Bayliss, P.~Eggenberger, J.-B.~Delisle, X.~Dumusque, M.~Marmier, R.~D\'\i az for discussions and insights about this work. We thank D.~Sosnowska, N.~Buchschacher, F.~Alesina for the help with the DACE platform.
\end{acknowledgements}


\bibliographystyle{aa}
\bibliography{hearts1}

\clearpage
\begin{appendix}

\section{MCMC results}\label{app:MCMC_DACE}
We show here the posterior probability distribution for the eccentricity obtained with the MCMC algorithm. This supports our choice to fix the eccentricity to zero.

\begin{figure}[htbp]
\centering
\includegraphics[width=0.47\textwidth]{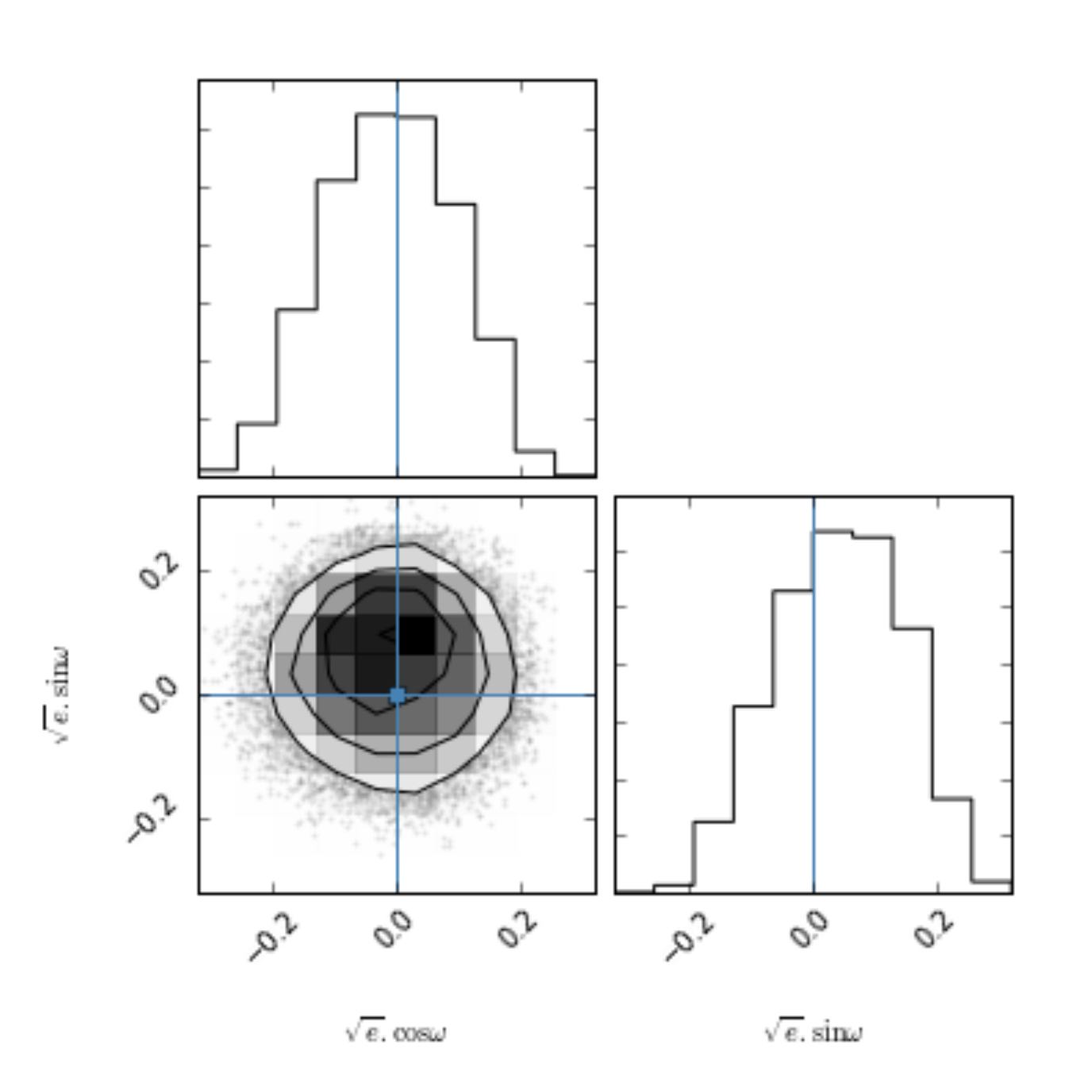}
\caption{Correlation diagrams for the probability distributions of $\sqrt{e}\mathrm{cos}\omega$ and $\sqrt{e}\mathrm{cos}\omega$. 1D histograms correspond to the distributions projected on the space of each parameter.}
\label{TS}
\end{figure}

\begin{figure}[htbp]
\centering
\includegraphics[width=0.47\textwidth]{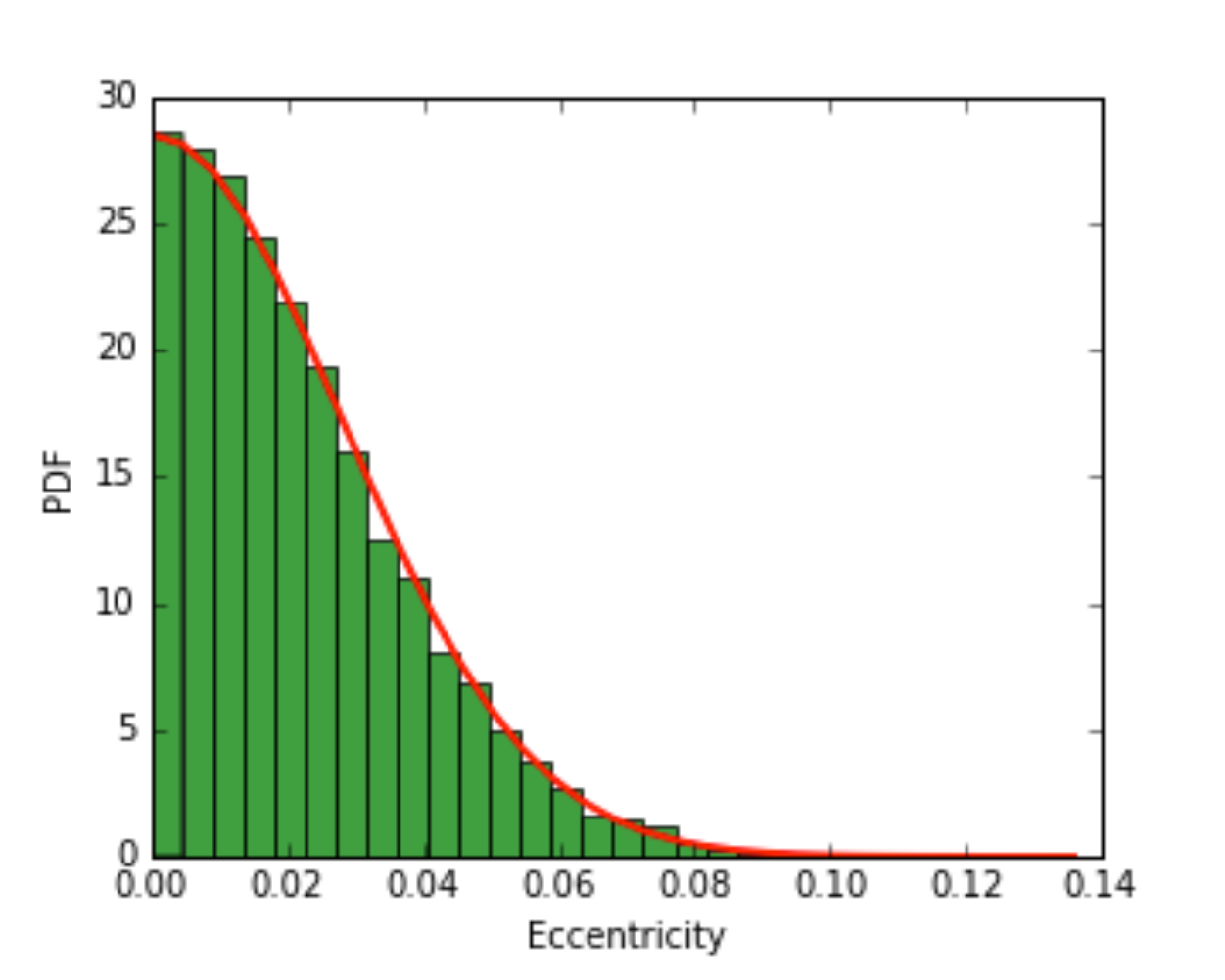}
\caption{Posterior probability distribution for the eccentricity.}
\label{TS}
\end{figure}

\section{On the ``transmission CCF''}\label{app:trans_ccf}

We compute the transmission CCF as follows:
\begin{equation}
\tilde{\mathfrak{R}}_{\mathrm{ccf}}(v) = \sum\limits_{t \in \mathrm{in}} \left. \frac{\widetilde{\mathrm{CCF}}(v,t)}{\widetilde{\mathrm{CCF}}_\mathrm{out}(v)}\right|_{p}.
\label{eq_CCF_TS1}
\end{equation}

Moreover, thanks to our measurement of the $\widetilde{\mathrm{CCF}}_{\mathrm{local}}$, the transmission CCF can also be simulated. To do so, we replace the numerator of Eq.~\ref{eq_CCF_TS1} by
\begin{equation}
\mathrm{CCF}(v,t) = \widetilde{\mathrm{CCF}}_{\mathrm{out}}(v) - [1-\delta(t)]\cdot\widetilde{\mathrm{CCF}}_{\mathrm{local}}(v,t)\ .
\label{eq_ccfin_simul}
\end{equation}
For the simulation we used the Gaussian fits to the master $\widetilde{\mathrm{CCF}}_{\mathrm{out}}$ and $\widetilde{\mathrm{CCF}}_{\mathrm{local}}$. Note the following subtlety: we had to shift the $\mathrm{CCF}_{\mathrm{res}}(t)$ (by $v_{\mathrm{surf}}(t)$) to construct our master $\widetilde{\mathrm{CCF}}_{\mathrm{local}}$. Then in order to simulate the observed $\mathrm{CCF}(v,t)$ in Eq.~\ref{eq_ccfin_simul}, we have to shift back the $\widetilde{\mathrm{CCF}}_{\mathrm{local}}$, i.e., $\widetilde{\mathrm{CCF}}_{\mathrm{local}}(v,t) = \widetilde{\mathrm{CCF}}_{\mathrm{local,+v_{surf}(t)}}(v)$.

\end{appendix}

\end{document}